\documentclass[12pt]{article}

\usepackage[margin=1in]{geometry}
\usepackage[T1]{fontenc}
\usepackage{mathpazo}
\usepackage{placeins}
\usepackage{setspace}
\usepackage{microtype}
\usepackage{amsmath}
\usepackage{array}
\usepackage{caption}

\usepackage{color}
\usepackage{colortbl}
\usepackage{enumitem}
\usepackage{graphicx}
\usepackage{multicol}
\usepackage{multirow}
\usepackage[table]{xcolor}
\usepackage[colorlinks=true, allcolors=blue]{hyperref}
\usepackage[longnamesfirst]{natbib}
\newcommand{\figalt}[1]{\par\smallskip\noindent\textit{Alt text:} #1}
\newcolumntype{P}[1]{>{\centering\arraybackslash}p{#1}}

\title{The Costs and Benefits of Leveraged ETFs\thanks{The authors would like to thank Benjamin Felix, Ivan Ivanov and Baolian Wang for helpful comments.}}

\author{
Chris Murray\thanks{Harvard Business School. \href{mailto:cdmurray80@gmail.com}{cdmurray80@gmail.com}}
\and
Marco Sammon\thanks{Harvard Business School. \href{mailto:mcsammon@gmail.com}{mcsammon@gmail.com}}
}

\date{October 31, 2026}

\begin{document}

\singlespacing

\setcounter{page}{0}

\maketitle
\thispagestyle{empty}

\begin{abstract}
\singlespacing

\noindent We study the costs and benefits of leveraged ETFs (LETFs), focusing on five determinants of performance: volatility drag, fees, financing costs, which products are launched, and market timing. Long broad-equity-index LETFs generated more than \$100 billion in investor gains, including over \$40 billion relative to counterfactual investments in the underlying assets, due to modest structural headwinds and rising markets. Single-stock LETFs, however, are launched on stocks near the top of the volatility and past-return distributions.  This caters to retail demand for lottery-like payoffs, but it also creates larger volatility drag, higher financing costs, and more risk for poor market-timing.


\end{abstract}

\newpage
\setcounter{page}{1}
\singlespacing

\section*{Introduction}

Even before SpaceX was publicly traded, ETF issuers were preparing products that would give investors daily levered exposure to its stock. While this example seems extreme, it actually reflects a broader shift in the leveraged ETF (LETF) market. Products that once focused mainly on broad equity indices now give investors levered exposure to individual stocks, cryptocurrencies, commodities, and volatility indices. Further, the LETF market has grown quickly in recent years: in 2025, leveraged long and inverse funds accounted for nearly one-fourth of new U.S. ETF launches, and leveraged-long ETFs reached a record \$122 billion in assets under management.

One reason LETFs have become popular is that they reduce the barriers to applying leverage to a given asset or strategy. Investors can obtain two or three times the \textit{daily} return of a benchmark without opening a margin account, trading derivatives, or rebalancing the position themselves. The fund sponsor embeds financing and daily rebalancing inside the ETF wrapper. This structure makes leverage more accessible, but it does not remove the costs of applying leverage to a given strategy. Over longer horizons, investor returns depend on the asset's cumulative return, the volatility of the asset's daily returns, management fees, financing costs, and how investors trade the product. These features matter most when leverage is applied to volatile individual companies, where compounding leveraged returns creates larger volatility drag and financing costs are higher on average.

A central question, therefore, is whether this easy access to leverage benefits investors. LETFs can be valuable when the asset they track has high realized returns, low volatility, and low financing costs. In fact, levered exposure to a diversified equity portfolio is consistent with standard life-cycle models in which young investors may want levered exposure to equities \citep{samuelson1969lifetime,merton1969lifetime,ayres2014practical}. The value added for investors can change, however, when the same structure is applied to volatile individual stocks.  In those products, volatility drag, fees, financing costs, and poorly timed flows can raise the breakeven return more than enough to offset the benefits of leverage, even when the underlying asset's return has been high.

In this paper, we document three sets of facts about the costs and benefits of LETFs. First, long broad-equity-index LETFs generated substantial realized gains for investors, including more than \$40 billion relative to counterfactual investments in the underlying assets. Second, single-stock LETFs are launched on unusually volatile recent winners, and products tied to more volatile stocks attract more AUM after launch. Third, LETF investors trade differently from investors in unlevered ETFs: flows are contrarian relative to recent returns, but do not predict subsequent performance, suggesting that investors in LETFs do not have market-timing skill. Together, these facts show that LETFs are not one product with one set of investor outcomes. Broad-equity-index LETFs have generally worked well for investors, while single-stock LETFs face much higher breakeven return hurdles.

We begin by measuring the gains and losses LETFs have generated for investors. Long LETFs in our sample generated roughly \$115 billion in investor gains, almost entirely from broad equity-index products. Specifically, we find that broad equity-index LETFs generated more than \$100 billion in gains, including more than \$40 billion relative to a counterfactual exercise in which investors placed the same flows at the same time into the corresponding unlevered assets. These results reject the standard story that volatility drag, fees, and financing costs \textit{always} dominate the benefits of leverage in LETFs.

We then show that the same daily leverage target produces very different multi-day returns depending on the volatility of the underlying asset. The key mechanism is \emph{volatility drag}: because LETFs compound leveraged daily returns, the cumulative log return of a long LETF is generally below the leverage multiple times the underlying asset's cumulative log return. And, this gap widens with realized volatility. We would like to emphasize, however, that although maintaining the daily leverage target requires each LETF to increase exposure after gains and reduce exposure after losses, volatility drag results from compounding leveraged returns rather than from the daily reset itself.

To illustrate the effect of volatility drag on long-run performance, we calculate the breakeven return -- defined as the return the underlying asset must achieve for the LETF and the unlevered underlying asset to yield the same cumulative return over a given year -- as a function of the underlying asset's volatility. For example, when the underlying asset's daily volatility is 5\%, a \(2\times\) LETF requires an annual underlying asset return of nearly 100\% to simply \textit{match} the performance of the unlevered asset.

Next, we estimate the other performance headwinds LETF investors face: expense ratios and funding spreads, i.e., the cost of borrowing above and beyond the risk-free rate to apply the stated leverage multiple. We find that expense ratios are close to 1\% for both equity-index and single-stock LETFs.  Funding spreads, however, differ significantly across broad equity-index and single-stock LETFs.  Specifically, broad equity-index LETFs have funding spreads below 1\% annually, while single-stock LETFs have estimated spreads between 6.8\% and 9.8\%, further raising the breakeven return for these products.

Next, we show that single-stock LETFs are launched precisely on stocks where these costs are likely to be high. Using only information available before launch, we rank the stocks underlying single-stock LETFs relative to the top 1,000 stocks by market capitalization. The median underlying stock is near the 94th percentile of the volatility distribution and the 92nd percentile of the past-six-month-return distribution. In other words, single-stock LETFs are launched on volatile recent winners. The nature of which LETFs get launched may reflect sponsors' efforts to cater to investor demand. Specifically, we find that among long single-stock LETFs, products tied to more volatile underlying stocks accumulate significantly more AUM, even after controlling for the underlying stock's market capitalization and fund age.

Another difference between LETFs on diversified portfolios and single stocks is how investors trade them. LETFs turn over much more rapidly than unlevered ETFs, with especially high turnover in single-stock, crypto, and VIX-linked products. Flows in LETFs also respond differently to past returns. Long LETF flows are negatively related to lagged one-day, one-week, and one-month returns, while unlevered ETF flows are on average positively related to recent returns. LETF investors therefore trade in a contrarian way, buying after recent losses and selling after recent gains. We find little evidence, however, that these flows predict future returns.  In other words, investors in LETFs do not appear to have market-timing skill.

Putting these facts together, we argue that whether LETFs benefit investors depends on which asset is being levered and how investors use the product. Fund sponsors have incentives to create products that attract assets, and retail investors are often drawn to assets with lottery-like payoffs \citep{kumar2009gambles,bali2011maxing,bali2023lotteryDemand}. Levered single-stock ETFs cater to this demand because they attach leverage to volatile individual stocks, increasing the perceived potential for large short-horizon gains. But the same volatility that makes the product seem attractive to investors also raises the breakeven return. Therefore, the products that look most attractive to ETF sponsors from an asset-gathering and trading-volume perspective can also be the products with the largest performance headwinds for end investors.

This distinction matters for how we evaluate LETFs as a financial innovation. Many debates about retail financial products focus on complexity, opacity, or hidden fees. LETFs are different, in that their daily objective is standardized across LETFs and easy to state. The harder question is whether the benefits of easy-to-trade levered exposure outweigh volatility drag, fees, and financing costs for the asset being tracked and the way investors ultimately trade the product. 

\subsection*{Literature Review}\label{sec:priorwork}

Our paper contributes to work on retail product design and how financial products cater to investors' preferences and biases. A common theme in this literature is that retail products emphasize salient features that attract demand while making performance headwinds less visible \citep{gabaix2006shrouded,bordalo2016competition}. This view is supported by a large body of work on structured products, for example research showing: (1) yield-enhancement products offer attractive stated yields but negative expected returns and are written disproportionately on volatile stocks \citep{vokata2021engineering} (2) issuers use complexity to reduce comparability across products and preserve markups \citep{chesney2026competition} and (3) products with higher headline rates tend to be more complex, riskier, and more profitable for banks \citep{celerier2017complexity,henderson2011dark}. LETFs differ because the daily objective is standardized and clearly stated. The less salient feature is how that daily objective translates into multi-day returns once volatility drag, fees, and financing costs are accounted for.

A second literature studies the return properties of LETFs themselves. \citet{cheng2009dynamics} emphasize that daily re-leveraging makes longer-horizon LETF returns path dependent and can make these products unsuitable for buy-and-hold investors. Relatedly, \citet{avellaneda2010path} show that LETF log returns can diverge from a simple multiple of the underlying asset's cumulative log return by a term linked to realized variance, and \citet{charupat2011pricing} examine the pricing and performance of leveraged ETFs. \citet{Wang2025MultiDay} studies the multi-day return properties of leveraged index ETFs and argues that LETFs do not generally have large performance headwinds, except when investors have relatively long holding periods in high-volatility assets. A related literature studies how LETF rebalancing affects the assets these products track. For example, \citet{ivanov2018leveraged} examine the effects of LETF rebalancing on underlying asset prices. 

More recent work studies single-stock LETFs. \citet{crouse2022averting} examines how volatility drag and tail risk are amplified when leverage is applied to individual stocks, and \citet{Bessembinder2025} decomposes the return shortfall of levered single-stock LETFs into daily rebalancing effects and frictions.  Our paper differs by not only studying the mechanics of LETFs, but more broadly assessing LETFs as \textit{investor products}. Our goal is to connect the mechanics of LETFs to the endogenous nature of product launches, the industry's shift from broad equity-index products to single-stock products, the specific products that attract assets, and how investors trade LETFs.

Our evidence on which LETFs get launched and attract assets connects to work on retail investor attention and demand for lottery-like assets. \citet{BarberOdean2008} show that individual investors buy attention-grabbing stocks, including stocks in the news, stocks with abnormal volume, and stocks with extreme returns. Relatedly, \citet{ben2023competition} show that ETF providers cater to investors' extrapolative beliefs by launching specialized ETFs that track underlying indexes with high recent past returns. \citet{kumar2009gambles} documents retail demand for lottery-type stocks, and \citet{bali2011maxing} show that stocks with extreme positive returns earn low subsequent returns. More recent work shows that retail trading is concentrated in hard-to-value stocks and salient retail names \citep{LaaritsSammon2025Retail,valeriaRobinHoodThisTimeDifferent}. We show that these forces matter not only for which products retail investors trade, but also for which levered products ETF sponsors create. 

Our results on LETF flows relate to work on contrarian trading by retail investors. \citet{BoehmerJonesZhang2021} show that retail investors trade individual stocks in a contrarian way and that retail order imbalance predicts future returns over short-horizons. \citet{kogan2024cryptos} document that trend-chasing versus contrarian trading can differ for the same investor across asset classes. We document a related contrast within the ETF market. Unlevered ETF flows respond positively to recent returns, while long LETF flows respond negatively to recent past returns.  However, LETF flows do not predict subsequent returns. This contrarian trading in LETFs, therefore, appears consistent with investors exhibiting the gambler's fallacy \citet{jinPeng2024SmallNumbers}. 

Finally, our paper contributes to the literature on retail leverage and retail options. Retail options are perhaps the closest substitute for levered single-stock LETFs, and recent work studies the motives, costs, and intermediation of retail option trading \citep{EatonGreenRosemanWu2026,bogousslavsky2024anatomy,ernst2026payment,bryzgalova2025strategic}. Our paper is also related to \citet{frazzini2022embedded}, who study securities with embedded leverage, including leveraged ETFs. Their paper, however, has a different focus: the \textit{price} of embedded leverage. Specifically, they show that when investors face leverage constraints, they may be willing to pay more for securities that provide levered exposure, lowering expected returns.

\section{Institutional Details and Data Sources}
\label{sec:mechanics_data}

This section describes how LETFs work and how we measure the main costs investors face when holding them: volatility drag, financing costs, and management fees. We then describe the LETFs that make up our sample and the data sources we use for our empirical analysis.  Finally, we provide summary statistics on LETF AUM by product category and how it has changed over time.

\subsection{How LETFs Work}
\label{sec:institutional_mechanics}

Leveraged exchange-traded funds (LETFs) are exchange-traded products that seek to deliver a fixed multiple of the \textit{daily} return of a stated benchmark.\footnote{For brevity, we use the term ``LETF'' to refer collectively to levered and inverse exchange traded funds (ETFs) and exchange traded notes (ETNs).} The benchmark may be an equity index, a commodity, a currency, a volatility index, a cryptocurrency, or a single stock. Let $R_{u(i),t}$ denote the daily return on the asset or benchmark tracked by product $i$, and let $L_i$ denote the product's target leverage multiple. The fund's daily return \textit{objective}, before fees, financing costs, trading costs, and tracking error, is:
\begin{equation}
  R^{LETF}_{i,t} \approx L_i R_{u(i),t}.
  \label{eq:daily_return_objective}
\end{equation}
For example, a $2\times$ LETF targets twice the benchmark's daily return, while a $-1\times$ inverse product targets the negative of the benchmark's daily return. The approximation in equation~\ref{eq:daily_return_objective} reflects the fact that realized daily returns for the LETF can differ from the stated objective due to management fees, financing costs, trading costs, and tracking error.

Most LETFs obtain leveraged exposure synthetically rather than by borrowing directly. A typical LETF uses total return swaps with bank counterparties, although some products also use futures, options, or other derivatives.\footnote{In a total return swap, the bank pays the fund the return on the benchmark for a specified dollar amount of exposure. In exchange, the fund pays the bank for the cost of providing that exposure, usually a short-term interest rate plus an additional charge -- i.e., the funding spread we measure in Section~\ref{sec:fees_financing} -- set by the bank. Importantly, these swaps allow the fund to obtain exposure struck at the official benchmark close, while the dealer hedges the residual exposure using market-on-close orders, futures, or other instruments. Direct futures trading by the LETF can create basis risk relative to the official benchmark close, which is used to evaluate the fund's tracking error. The swap financing costs are typically not reported, so we estimate them in Section~\ref{sec:fees_financing} based on the gap between the fund's realized returns and the returns implied by its daily leverage objective and stated management fees.} While the exact implementation varies across products, the broad idea is the same across all LETFs: the leverage is embedded inside the ETF wrapper, so the investor does not need to borrow, trade derivatives, or rebalance the position directly.

Matching a daily leverage target requires the LETF to rebalance every day. For example, consider a $2\times$ LETF with \$50 of investor equity and \$100 of exposure to the benchmark, with the extra \$50 obtained through synthetic exposure. If the benchmark rises by 10\%, the position is worth \$110 and the investor equity rises from \$50 to \$60. The LETF is now levered only \$110/\$60 = $1.83\times$. To get back to $2\times$, the LETF must re-lever, i.e., increase exposure. The same logic works in reverse after the benchmark falls: equity falls, leverage rises above the target, and the LETF must de-lever, i.e., reduce exposure. As we explain in the next subsection, this daily rebalancing is why multi-day LETF returns depend on the path of daily returns, not just the benchmark's cumulative return.

\subsection{Volatility Drag and Breakeven Returns}
\label{sec:volatility_drag}

In a frictionless market, the cumulative gross return of an LETF over $N$ trading days is
\begin{equation}
  1 + R^{LETF}_{0:N}
  =
  \prod_{t=1}^{N} \left(1 + L R_{u,t}\right),
  \label{eq:letf_compound_return}
\end{equation}
where $L$ is the LETF's leverage multiple and $R_{u,t}$ is the return on the underlying asset.
Using a second-order approximation, the difference between the LETF's cumulative log return and $L$ times the underlying asset's cumulative log return is approximately
\begin{equation}
  \sum_{t=1}^{N} \log\left(1 + L R_{u,t}\right)
  -
  L \sum_{t=1}^{N} \log\left(1 + R_{u,t}\right)
  \approx
  -\frac{1}{2} L(L-1) \sum_{t=1}^{N} R_{u,t}^{2}.
  \label{eq:vol_drag_realized}
\end{equation}
Equivalently, if the underlying asset has per-period volatility $\sigma_u$, the effect of volatility on cumulative returns is approximately
\begin{equation}
  DRAG_{RET}
  =
  \exp\left(
    - \frac{N L(L-1)}{2}\sigma_u^2
  \right).
  \label{eq:lev_drag_ret}
\end{equation}
Equation~\ref{eq:lev_drag_ret} shows how realized variance affects longer-horizon (i.e., greater than 1-day) LETF returns. For long LETFs with $L>1$, higher realized variance lowers the LETF's cumulative log return relative to $L$ times the underlying asset's cumulative log return \citep{avellaneda2010path,cheng2009dynamics}.\footnote{The intuition behind Equation~\ref{eq:lev_drag_ret} is easiest to see with a two-day reversal example. Suppose that the underlying asset falls by 1\% one day and then rises by 1\% the following day. A $2\times$ LETF earns $-2\%$ and then $+2\%$, for a two-day return of $(0.98)(1.02)-1=-0.04\%$. With larger daily swings, the cumulative loss grows quickly. For example, if the underlying asset falls by 5\% and then rises by 5\%, the $2\times$ LETF earns $-10\%$ and then $+10\%$, for a two-day return of $(0.90)(1.10)-1=-1.00\%$.} The same logic applies to inverse products: when $L<0$, the term $L(L-1)$ is also positive.

The variance term $\sigma_u^2$ in Equation~\ref{eq:lev_drag_ret} shows that the gap between the LETF's log return and $L$ times the asset's log return grows with the \textit{square} of realized volatility. This gap can therefore be much larger for single stocks than for broad indices. In 2025, for example, the AUM-weighted daily volatility of assets tracked by leveraged single-stock ETFs was 3.71\%, compared with 1.35\% for leveraged equity-index ETFs. Holding leverage fixed, this difference increases the volatility-drag term by a factor of approximately $(3.71/1.35)^2=7.6 \times$.

Figure~\ref{fig:realized_drag_2x} provides fund-level evidence on this mechanism for $2\times$ long LETFs. The figure plots each fund's annualized log-return gap on the y-axis, defined as the fund's annualized log return minus two times the annualized log return of its underlying asset, against the underlying asset's realized daily volatility on the x-axis. The realized gaps (the blue dots) become more negative as underlying volatility rises and broadly follow the frictionless volatility-drag prediction from Equation~\ref{eq:lev_drag_ret} (the red line).

To quantify this effect, we use equation~\ref{eq:lev_drag_ret} to compute a breakeven return: the annual return the underlying asset must earn for a long LETF to \textit{match} the unlevered asset over one year given $L$ and $\sigma_u$. Panel A of Table~\ref{tab:breakeven_vol_drag} reports this breakeven return for $2\times$ and $3\times$ leveraged funds under a range of daily volatility assumptions. Table~\ref{tab:breakeven_vol_drag} shows that this hurdle rises quickly with volatility. For a $2\times$ LETF, the breakeven return is 8.4\% when daily volatility is 1\%, 16.9\% when daily volatility is 2\%, and 98.4\% when daily volatility is 5\%. In other words, when daily volatility is 5\%, the underlying asset must earn nearly 100\% over the year for the $2\times$ LETF to match an unlevered position in the same asset.

At first glance, this seems surprising. As long as the underlying asset has a positive annual return, one might expect the $2\times$ long product to \textit{always} outperform the unlevered asset.
Equation~\ref{eq:lev_drag_ret}, however, shows why that need not be true. When daily volatility is high, volatility drag can offset the benefit of leverage unless the underlying asset's cumulative return is very large. Table~\ref{tab:breakeven_vol_drag} also shows that for a $3\times$ product, the breakeven return rises even faster with the volatility of the underlying asset. Appendix Section~\ref{app:vol_drag_details} provides numerical examples and details on our volatility drag calculation.

\subsection{Fees and Financing Costs}
\label{sec:fees_financing}

While volatility drag is specific to leveraged products, LETF investors also face management fees and financing costs. Management fees are familiar from unlevered ETFs and are disclosed directly through expense ratios. Financing costs reflect the cost of maintaining the fund's stated leveraged exposure. Because these costs are built into swap, futures, options, or other derivative pricing, we infer them from realized LETF returns.

Panel A of Table~\ref{tab:costs} reports asset-weighted expense ratios for the LETFs in our sample. Expense ratios are generally close to 1\% per year for equity-index and single-stock LETFs, and higher for some other categories, such as crypto and volatility-linked products. These fees are much higher than the fees charged by large unlevered index ETFs, which are typically in the range of 3--5 bps per year.

We estimate financing costs by assuming that the fund pays the risk-free rate plus a funding spread on the extra exposure needed to reach its leverage target. For example, consider a $2\times$ long ETF with \$1 million in investor capital. To deliver twice the benchmark's daily return, the fund needs approximately \$2 million of total benchmark exposure. We treat the additional \$1 million of exposure as being financed at the risk-free rate plus an (unknown) spread, and then estimate that spread using the gap between the fund's realized return and the return implied by its daily leverage objective and management fees. For a long product with $L_i>1$, the daily return can be written as
\begin{equation}
  R^{LETF}_{i,t}
  =
  L_i R_{u(i),t}
  -
  (L_i-1)\left(R_{rf,t}+\frac{s_i}{252}\right)
  -
  \frac{ER_i}{252}
  +
  \epsilon_{i,t},
  \label{eq:letfDailyPerformance}
\end{equation}
where $R_{rf,t}$ is the daily risk-free rate, $s_i$ is the annualized net funding spread on the incremental leveraged exposure, $ER_i$ is the annual expense ratio, and $\epsilon_{i,t}$ captures tracking error and other residual components. Throughout the paper, we assume that there are 252 trading days per year.

To estimate these funding spreads empirically, we first define the ideal daily return net of the risk-free financing component and expense ratio as
\begin{equation}
  R_{i,t}^{ideal}
  =
  L_i R_{u(i),t}
  -
  (L_i-1)R_{rf,t}
  -
  \frac{ER_i}{252}.
  \label{eq:ideal_return}
\end{equation}
We then estimate the daily return gap between the LETF and this ideal return as
\begin{equation}
  R^{LETF}_{i,t} - R_{i,t}^{ideal}
  =
  -\beta_1 \frac{L_i-1}{252}
  +
  \epsilon_{i,t}.
  \label{eq:funding_regression}
\end{equation}
In equation~\ref{eq:funding_regression}, $\beta_1$ is the estimated annualized funding spread. The negative sign reflects the fact that funding costs reduce realized returns relative to the ideal return. We scale the $\beta_1$ term by $L_i-1$ because the funding spread applies only to the \textit{incremental} exposure financed by the fund.

Panel B of Table~\ref{tab:costs} summarizes our estimated funding spreads. We find that equity-index LETFs have estimated funding spreads below 1\% annually. On the other hand, estimated funding spreads for single-stock LETFs are much larger, at between 6.8\% and 9.8\% for $2\times$ products. These higher spreads matter because they are paid on the \textit{entire} additional exposure the fund uses to maintain its leverage target. For a $2\times$ product, the fund needs one extra dollar of exposure for each dollar of investor capital. For a $3\times$ product, it needs two extra dollars of exposure for each dollar of investor capital, so even with the same funding spread, the performance drag would be double.

Fees and financing costs create an additional annual return drag, separate from the volatility drag discussed in Section~\ref{sec:volatility_drag}. For a $2\times$ long LETF, each dollar of investor capital requires one additional dollar of exposure. Investors therefore effectively pay the risk-free rate and funding spread on that additional exposure, plus the fund's expense ratio. Using the end-of-sample risk-free rate and the estimates from Table~\ref{tab:costs}, this annual cost is approximately $4.09\% + 0.4\% + 0.90\% = 5.39\%$ for a $2\times$ equity-index LETF. For a $2\times$ single-stock LETF, the analogous cost is approximately $4.09\% + 6.8\% + 1.03\% = 11.92\%$.

\subsection{Data and Sample Selection}
\label{sec:data}

Our main LETF sample consists of long leveraged and inverse exchange-traded funds and exchange-traded notes from 2006 through September 2025. We identify the initial universe using the CRSP Mutual Fund database, Bloomberg classifications, fund names, fund prospectuses, and other public disclosures. We then manually screen funds to verify that they have levered or inverse mandates.

From this universe, we apply two filters. First, we exclude LETFs with monthly or weekly rebalancing frequencies because their return mechanics differ from the daily-reset LETFs discussed in Section~\ref{sec:volatility_drag}. We also exclude LETFs that use options as the primary way to target leveraged exposure, because option-based products have payoff nonlinearities that are distinct from the LETF return mechanics we are focused on. Second, to reduce the influence of very small and illiquid funds, we require that a fund reach at least \$30 million in assets under management in at least one calendar month during its life.  

We use public disclosures to identify each LETF's leverage ratio and the asset or benchmark it tracks.\footnote{To validate these mappings, we use daily returns. Specifically, for each fund, we regress the LETF's daily return on the return of the tracked asset multiplied by the fund's leverage ratio. If the mapping is correct, this should explain most of the fund's daily return variation, because the product's stated objective is a multiple of daily returns before fees, financing costs, and tracking error. We keep products without a clean return mapping in the aggregate AUM and flow analyses, but exclude them from tests that require daily returns on the asset the product tracks.} Some funds change their leverage ratio over time, so we record the time-varying leverage ratio when this occurs. For single-stock LETFs, the tracked asset is the referenced individual stock. For index LETFs, we map the product to the stated index or to an unlevered ETF that tracks the same benchmark when such a product is available. Using an unlevered ETF as the return proxy gives us a tradable measure of benchmark returns that includes ordinary ETF trading and tracking frictions, rather than relying on an index return that investors may not be able to buy directly.

We obtain return data from CRSP daily stock data. Because CRSP typically updates ETF shares outstanding monthly rather than daily, we merge in daily shares outstanding data from Bloomberg, which allow us to  measure creations and redemptions (i.e., flows) at the daily frequency. We define assets under management as shares outstanding multiplied by price.

Following the literature on flow-induced trading by ETFs \citep{schmickler2020identifying}, we estimate daily net dollar flow as the split-adjusted daily change in shares outstanding multiplied by the daily closing price:
\begin{equation}
  Flow_{i,t}
  =
  \Delta Shares_{i,t}
  \times
  Price_{i,t}.
  \label{eq:flow}
\end{equation}
Positive values correspond to net creations (fund-level inflows) while negative values correspond to net redemptions (fund-level outflows). In calculating net flows, we assume creations and redemptions occur at the closing price.\footnote{The measure of flows in Equation \ref{eq:flow} captures net fund-level flow, not secondary-market trading. If one investor sells ETF shares to another investor on an exchange, shares outstanding do not change and our flow measure is zero. By contrast, if aggregate demand leads authorized participants to create new ETF shares, shares outstanding increase and our measure will treat this as an inflow. }

We use the same data from CRSP and Bloomberg to construct the unlevered ETF comparison sample. The comparison sample consists of approximately 50 of the largest unlevered ETFs by assets under management. These products allow us to compare flows and trading behavior in LETFs to flows and trading behavior in standard ETFs using the same measurement approach in Section~\ref{sec:flow_timing}.

\subsection{Summary Statistics}
\label{sec:summary_statistics}

Figure~\ref{fig:aum_by_etf_type} reports assets under management by LETF type, and Figure~\ref{fig:flow_by_etf_type} reports cumulative investor flows using equation~\ref{eq:flow}. Total AUM and cumulative net flows are related, but they are not the same object. Net flows just measure creations and redemptions. AUM also reflects when products enter the sample through new launches and the return on capital of existing funds. 

Both measures are different from secondary-market trading volume, which measures how often ETF shares change hands between investors.  To quantify this, we also summarize turnover by product type in Table~\ref{tab:turnover_by_period_type}. Panel A reports the number of ETFs in each category by decade. Panel B reports total daily turnover, defined as aggregate dollar volume divided by aggregate AUM within each category, and Panel C reports the mean of product-level daily turnover. Net flows show whether aggregate exposure to a product is rising or falling through creations and redemptions. Turnover shows how heavily investors trade the product even when shares outstanding do not change. Table~\ref{tab:turnover_by_period_type} shows that broad equity-index LETFs have historically accounted for much of the LETF market, while single-stock and crypto LETFs were launched mainly in the last few years. These newer products (along with volatility-linked products) have much higher turnover than standard unlevered ETFs.

\section{Product Entry and Catering to Investor Demand}
\label{sec:entry_demand}

We next study which LETFs sponsors choose to create and which products attract investor flows after launch. This matters because the products that attract attention, AUM, and trading volume are not necessarily the products that will deliver the highest expected returns for investors. Specifically, we show that single-stock LETFs are launched on volatile recent winners, that products tied to more volatile stocks attract more AUM, and that these products also face the highest performance headwinds.

\subsection{From Indexes to Single Stocks}
\label{subsec:evolution_letf_market}

When they were first launched, LETFs mainly offered investors exposure to broad equity indices, sectors, commodities, currencies, bonds, and volatility indices. Many of these products, especially broad equity-index LETFs, track diversified portfolios with relatively low daily volatility.\footnote{Volatility-linked products are the main exception because their underlying benchmarks, typically VIX futures, are much more volatile. As a specific example, XIV, an inverse VIX ETN, lost over 90\% of its value in February 2018 and was subsequently closed.} Single-stock LETFs apply the same daily leverage multiple to individual companies. This moves leverage from diversified portfolios to more volatile underlying assets, where volatility drag and financing costs are larger on average.

\begin{figure}[!htbp]\centering
\includegraphics[
height=3.0in,
alt={Two line charts show assets under management by leveraged ETF type over time. Leveraged equity-index ETFs dominate total assets for most of the sample and rise sharply after 2020. When leveraged equity-index ETFs are excluded, single-stock LETFs become the largest and fastest-growing category in the 2020s, with crypto products also increasing near the end of the sample.}
]{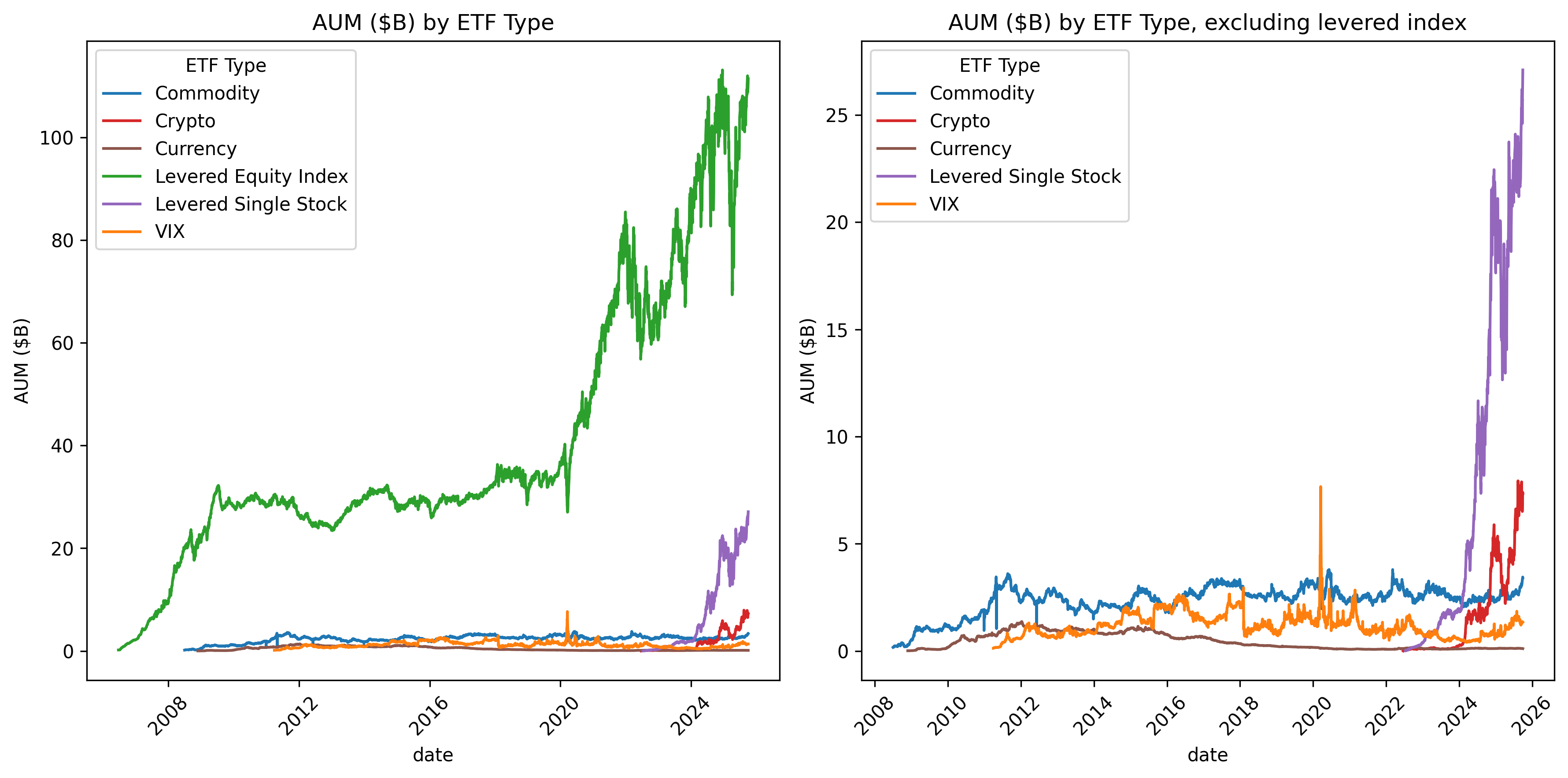}
\caption[]{\textbf{AUM by LETF Type.} The figure reports total assets under management by LETF type. The left panel includes all LETFs in the sample, while the right panel excludes leveraged equity-index ETFs.}
\figalt{Two line charts show assets under management by leveraged ETF type over time. Leveraged equity-index ETFs dominate total assets for most of the sample and rise sharply after 2020. When leveraged equity-index ETFs are excluded, single-stock LETFs become the largest and fastest-growing category in the 2020s, with crypto products also increasing near the end of the sample.}
\label{fig:aum_by_etf_type}
\end{figure}

In our sample, single-stock LETFs were first launched in 2022, with products tied to large, actively traded stocks such as Tesla, NVIDIA, PayPal, Nike, and Pfizer. As the market expanded, issuers began to launch products tied to stocks with significant interest by retail investors (i.e., ``meme stocks''), such as MicroStrategy. Figures~\ref{fig:aum_by_etf_type} and~\ref{fig:flow_by_etf_type} show how this entry changed the LETF market. Before 2023, leveraged equity-index ETFs accounted for most LETF assets and flows relative to commodity, VIX, and currency LETFs. Beginning in 2023, single-stock LETFs began to attract large inflows and became an increasingly important part of the overall LETF market.

\subsection{Which Stocks Get LETFs?}
\label{subsec:which_stocks_get_letfs}

We characterize the stocks that receive single-stock LETFs by ranking each underlying stock on realized volatility and past-six-month return among the top 1,000 ordinary common shares traded on major exchanges in CRSP by market capitalization, using only information available before the launch date. Because this analysis uses the single-stock LETFs \textit{in our sample}, it is limited to launches that attracted enough assets to satisfy the AUM threshold in Section~\ref{sec:data}.

Panel A of Table~\ref{tab:single_stock_launch_aum} reports percentile ranks for realized volatility and past-six-month returns. We show that LETF launches are concentrated in the very top of both distributions. The median underlying stock is near the 94th percentile of the volatility distribution, and 86\% of launches occur on stocks above the 75th percentile. Recent returns show a similar pattern: the median launch stock is near the 92nd percentile of past-six-month returns, and more than 62\% of launches occur on stocks above the 75th percentile.  

In summary, Table~\ref{tab:single_stock_launch_aum} shows that single-stock LETFs are launched on volatile recent winners. This is our first piece of evidence that product entry in the LETF space likely caters to investor demand. Specifically, launching LETFs on volatile recent winners is consistent with work showing that retail investors are drawn to salient, lottery-like stocks. Recent winners are more likely to attract attention \citep{BarberOdean2008}, and volatile stocks offer larger potential short-horizon gains. Adding a leverage multiple on top of these extreme recent returns likely makes these LETFs even more attention-grabbing. This also relates to work measuring lottery demand using stocks' recent extreme positive returns, which shows that these stocks attract investor demand \citep{kumar2009gambles,bali2011maxing,bali2017lotteryBeta,bali2023lotteryDemand}. We next ask whether the same characteristics that predict product entry also predict which single-stock LETFs attract assets after launch.

\subsection{Which Products Attract Assets?}
\label{subsec:single_stock_letf_assets}

LETF sponsors earn fees on assets under management, so they have incentives to launch products they expect will gather assets. The next question is therefore which products investors allocate flows to after launch. We find that broadly, the evidence on flows lines up with the product-entry results in the previous subsection: single-stock LETFs tied to more volatile stocks attract more AUM.

Panel B of Table~\ref{tab:single_stock_launch_aum} reports fund-level regressions of log average daily AUM on the underlying stock's average daily volatility and log market capitalization. We control for market capitalization because larger companies may naturally support larger ETF products, may be more salient to investors, and because high-volatility stocks tend to be smaller. Column (1) includes all long leveraged single-stock ETFs with non-missing underlying data and controls for fund age using the log number of trading days. Column (2) restricts the sample to funds with at least 252 trading days in our dataset.

Our estimated effect of underlying volatility on AUM is large. In column (1), a one percentage point increase in underlying daily volatility is associated with a 0.52 log-point increase in average AUM, or approximately 68\% higher average AUM ($e^{0.52}-1 \approx 0.68$), controlling for stock market capitalization and fund age. In column (2), among funds with at least 252 trading days, the corresponding estimate is 0.77 log points, or roughly 116\% higher average AUM.  As a robustness check, Appendix Table~\ref{tab:app_rank_aum_regressions_vs_vol} performs a similar regression using the rank of AUM and volatility, with similar results.

\subsection{Trading Volume}
\label{subsec:which_products_attract_trading}

AUM matters because sponsors earn fees on assets, but trading volume also matters for product success. Heavily traded products are more visible in product rankings, more liquid, and therefore may become easier for investors to find and trade. If LETFs were mainly long-horizon asset-allocation tools, secondary-market turnover would be low. If they cater to investors who want to trade salient, speculative exposures, turnover may be higher, especially in single-stock, crypto, and volatility-linked LETFs.

Table~\ref{tab:turnover_by_period_type} reports ETF counts and turnover by product type and decade. Panel B reports total daily turnover, defined as aggregate dollar volume divided by aggregate AUM within a product category. Panel C reports the mean of product-level daily turnover. We also split unlevered ETFs into high- and low-volume groups so that the comparison is not driven by comparing LETFs only to low-volume buy-and-hold products.

We find that LETFs trade much more than unlevered ETFs. In the 2020s, leveraged equity-index ETFs have total daily turnover of 16.37\% and mean turnover of 8.34\%, compared with 5.51\% and 5.83\% for the highest-volume unlevered ETFs. The gap is larger in the newer, more speculative categories. For example, leveraged single-stock ETFs have total daily turnover of 35.08\% and mean turnover of 44.26\%. Crypto LETFs have total daily turnover of 20.86\%, and VIX products have mean turnover above 75\%.

\subsection{Retail Trading in LETFs}
\label{subsec:retail_trading_letfs}

The product-entry and turnover results raise a natural question: who trades these products, and therefore whose demand are LETF sponsors catering to? Retail investors are the group most often associated with demand for salient, volatile, lottery-like assets, and single-stock LETFs offer exactly these kinds of exposures. To test this, we use data from VandaTrack, which applies a proprietary algorithm to order flow data to identify retail-initiated buy and sell trades.\footnote{Importantly, this measure does not represent total retail-initiated volume, as VandaTrack only captures a fraction of the retail market.  However, it is still useful for comparing the intensity of retail trading across different groups of securities \citep{LaaritsSammon2025Retail}. As with any trade-classification algorithm (e.g., the retail-initiated trade identification method in \citet{BoehmerJonesZhang2021}), the level should not be interpreted as the true fraction of all trading done by retail investors. Prior work shows that retail-trade identification algorithms can miss many true retail trades and can also misclassify some trades as retail which are initiated by institutions \citep{barber2024sub,battalio2023difficulties}. We therefore focus on the relative ranking across products and stocks, not the absolute level of the retail share of trading volume.}

We start by sorting securities into several groups: leveraged ETFs on broad indices, leveraged ETFs on individual stocks, the matched set of broad index ETFs, the matched set of individual stocks, and a set of control stocks (ordinary common shares traded on major exchanges) with no corresponding leveraged ETF.  For each group, we calculate the retail share of volume by summing dollars of retail-initiated buys and sells (i.e., gross retail order flow) and dividing by the total dollar trading volume across all securities in that group. By summing dollars of trading volume across all securities in the group, these metrics are effectively value-weighted.

\begin{figure}[t]
\centering
\includegraphics[
width=\linewidth,
alt={Two line charts show 22-day moving averages of the share of dollar trading volume attributed to retail investors. The left panel shows that leveraged equity-index ETFs have persistently higher retail trading shares than unlevered ETFs. The right panel shows that stocks underlying single-stock LETFs have higher retail trading shares than other stocks, and that levered single-stock ETFs themselves have especially high retail trading shares after launch.}
]{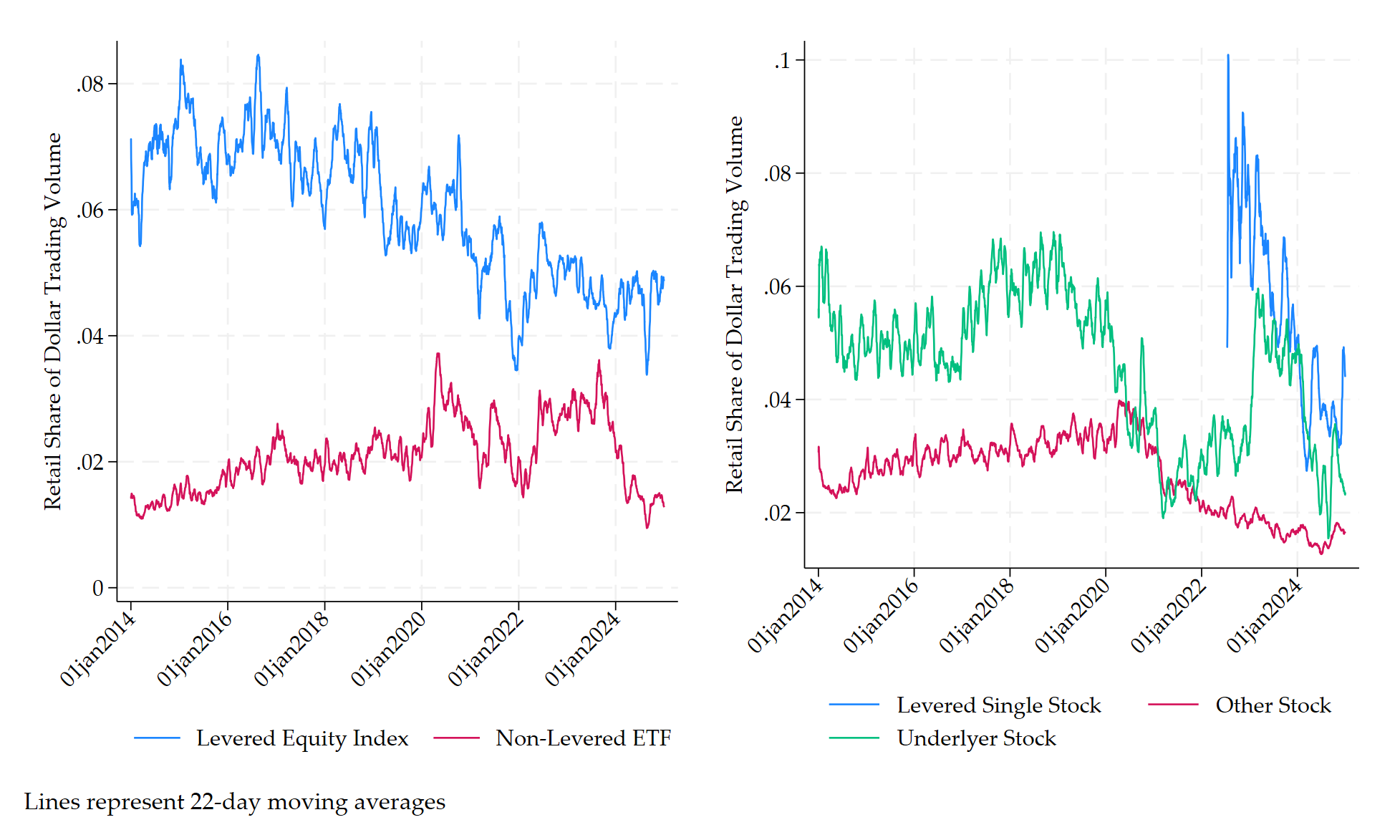}
\caption[]{\textbf{Retail-Attributed Trading in LETFs and Their Underlying Stocks.} This figure plots 22-day moving averages of retail-attributed dollar trading volume divided by total dollar trading volume using VandaTrack retail order-flow data. The measure is constructed at the group-day level by summing retail-attributed buys and sells across all securities in each group and dividing by total dollar trading volume across those same securities. It is therefore dollar-weighted across securities within each group. The left panel compares leveraged equity-index ETFs to unlevered ETFs. The right panel compares levered single-stock ETFs, the stocks underlying those ETFs, and other stocks. The measure should be interpreted as a proxy for retail trading intensity rather than the true level of all retail trading.}
\figalt{Two line charts show 22-day moving averages of the share of dollar trading volume attributed to retail investors. The left panel shows that leveraged equity-index ETFs have persistently higher retail trading shares than unlevered ETFs. The right panel shows that stocks underlying single-stock LETFs have higher retail trading shares than other stocks, and that levered single-stock ETFs themselves have especially high retail trading shares after launch.}
\label{fig:retail_share_letfs}
\end{figure}

Figure~\ref{fig:retail_share_letfs} plots 22-day moving averages of this measure of the retail share of trading volume. The left panel compares leveraged equity-index ETFs to unlevered ETFs. Leveraged equity-index ETFs have a much higher retail-attributed share of dollar trading volume throughout the sample. The right panel compares three groups: levered single-stock ETFs, the stocks underlying those ETFs, and other stocks. The stocks that receive single-stock LETFs have higher retail-attributed trading shares than other stocks, even before the single-stock LETFs are launched. After launch, the levered single-stock ETFs themselves have especially high retail-attributed trading shares.

These patterns support the interpretation that LETFs are catering to retail investor demand. Sponsors launch single-stock LETFs on volatile recent winners -- which are typically attractive to retail investors.  Figure~\ref{fig:retail_share_letfs} directly shows that the products sponsors create, and the stocks on which they create them, are also disproportionately favored by retail traders.

\section{Investor Trading and Market Timing}
\label{sec:flow_timing}

We next study when investors add and withdraw capital from LETFs. Because LETFs may attract a different clientele than ordinary unlevered ETFs, their flows may respond differently to past returns. We show that LETF investors allocate more capital after recent losses and withdraw capital after recent gains. These contrarian flows differ from the return-chasing flows in unlevered ETFs, but they do not predict future returns i.e., we find no evidence of market-timing skill by LETF investors.

\subsection{Relationship of Flows to Past Returns}
\label{subsec:buying_dips}

We begin by estimating how daily flows in LETFs respond to past returns. For each long LETF\footnote{In this subsection, we focus on long LETFs because inverse products have a different mechanical interpretation. For an inverse product, a loss on the fund corresponds to a gain in the asset it tracks, which complicates the interpretation of buying after gains or losses.}, we regress normalized daily flows on lagged fund returns over multiple horizons:
\begin{equation}
\begin{aligned}
\widetilde{Flow}_{i,t}
&=
\alpha
+
\beta_1 R_{i,t-1}
+
\beta_5 R_{i,t-5:t-1}
+
\beta_{21} R_{i,t-21:t-1}
\\
&\quad+
\beta_{63} R_{i,t-63:t-1}
+
\beta_{126} R_{i,t-126:t-1}
+
\epsilon_{i,t}.
\end{aligned}
\label{eq:flow_past_returns}
\end{equation}
To make flows comparable across funds, we winsorize daily flows at the fund level and then standardize them by subtracting the fund-level mean and dividing by the fund-level standard deviation. We winsorize before standardizing so that a small number of unusually large creations or redemptions do not drive the estimates.

Panel A of Table~\ref{tab:flow_timing_regressions} reports the results, showing that LETF flows are strongly contrarian. The coefficients on lagged 1-day, 5-day, and 21-day returns are negative and statistically significant in the full sample and in each subperiod. Further, we find that the response of flows to past performance is strongest at short horizons. In the full sample, the coefficient on the lagged 1-day return is $-4.02$, and the coefficient on the lagged 5-day return is $-2.59$. This behavior is consistent with past work showing that retail investors trade in a contrarian way relative to recent returns in other settings \citep{BoehmerJonesZhang2021,kogan2024cryptos,luo2025retail,valeriaRobinHoodThisTimeDifferent}.

The pattern we document in LETFs differs significantly from the flow-performance relationship for unlevered ETFs. Panel B of Table~\ref{tab:flow_timing_regressions} repeats the same analysis for the unlevered ETF comparison sample, showing that unlevered ETF flows respond positively to recent returns. In the full sample, the coefficient on the lagged 1-day return is $2.23$, and the coefficient on the lagged 5-day return is $2.42$. Flows into standard ETFs therefore look more like return chasing, whereas flows into long LETFs look like short-horizon reversal trading.

\subsection{Measuring Market-Timing Skill}
\label{subsec:flows_future_returns}

Buying after recent losses is not necessarily good or bad. It can improve returns if recent losses are temporary price declines that subsequently reverse \citep{nagel2012evaporating}. It can hurt returns if recent losses reflect information and returns continue in the same direction \citep{novy2015fundamentally}. We therefore test whether LETF flows contain market-timing skill by asking whether inflows predict future fund returns. Specifically, we augment the flow regression in equation~\ref{eq:flow_past_returns} with future fund returns:
\begin{equation}
\widetilde{Flow}_{i,t}
=
\alpha
+
\sum_{h \in H} \beta_h R_{i,t-h:t-1}
+
\sum_{h \in H} \delta_h R_{i,t+1:t+h}
+
\epsilon_{i,t},
\label{eq:flow_future_returns}
\end{equation}
where $H=\{1,5,21,63,126\}$.  A positive coefficient on future returns would mean that inflows arrive before strong subsequent performance, conditional on the past returns that preceded the flows. This is our test of market-timing skill.

Table~\ref{tab:flowTimingRegressionFutureletf1} reports the results. The coefficients on lagged returns remain large, negative, and statistically significant, consistent with Table~\ref{tab:flow_timing_regressions}. The coefficients on future returns, however, show no stable positive pattern across horizons or subperiods. Many estimates are negative, and the positive estimates appear inconsistently. These results suggest that LETF investors buy after losses, but do not identify which losses will reverse. This behavior is consistent with investors expecting short-term reversals, as in the gambler's-fallacy mechanism studied by \citet{jinPeng2024SmallNumbers}. In our data, however, LETF flows do not predict future returns.  We interpret this as evidence that LETF investors lack market-timing skill.

Past returns predict LETF flows, but they do not by themselves tell us what returns investors actually earned in LETFs. Timing matters because a fund can perform well on average and still deliver poor investor returns if most capital arrives before large losses. XIV, an inverse VIX ETN, is the extreme example of this: strong prior performance attracted significant AUM, but the product lost most of its value after the February 2018 volatility spike and eventually closed. Investors who entered near the end of XIV's life therefore experienced losses that would not be obvious from the simple average of XIV's daily returns.

The next section brings these pieces together. We ask whether LETFs generated dollar gains or losses for investors, how those outcomes compare to investments in the underlying assets, and how much of investor performance reflects structural product headwinds, fund selection, and within-fund market timing.

\section{Investor Outcomes}
\label{sec:investor_outcomes}

Investor outcomes in LETFs depend on the forces established in the previous sections: the return on the asset being levered, volatility drag, fees, financing costs, which products sponsors launch, and when investors allocate capital. This section translates those forces into realized dollar outcomes for LETF investors. We start by measuring aggregate investor gains and losses. We then compare those outcomes to investments in the assets the products track. Finally, we decompose investor returns into product returns, fund selection, and within-fund market timing.

\subsection{Aggregate Gains and Losses}
\label{subsec:aggregate_investor_gains_losses}

Our baseline measure of investor performance is aggregate realized profit and loss, which we call \textit{Investor PnL}. For fund $i$ on day $t$, daily Investor PnL is beginning-of-day assets under management multiplied by the fund's daily return.\footnote{Because this measure uses beginning-of-day assets under management and realized daily returns, it does not include intraday secondary-market execution costs paid by investors.} We then sum across funds and days:
\begin{equation}
  \text{Investor PnL}
  =
  \sum_i \sum_t AUM_{i,t-1} R_{i,t}.
  \label{eq:investorpnl}
\end{equation}
This measure weights returns by the dollars invested when those returns are realized. As discussed in the previous subsection, large returns before assets arrive matter little for aggregate investor outcomes, while drawdowns after large inflows can generate large dollar losses.

Panel A of Table~\ref{tab:investor_pnl_combined} reports aggregate investor PnL, net flows, and end-of-sample assets under management. Long LETFs generated \$114.8 billion in investor PnL on \$22.8 billion of net flows. Inverse LETFs look very different: they attracted \$93.4 billion in net flows but generated investor losses of \$80.3 billion. While these losses are significant, we believe they should be interpreted cautiously. Inverse LETFs may be used to hedge other exposures, so inverse LETF losses may be offset by gains elsewhere in investors' portfolios.  Further, our sample includes unusually strong equity-market returns, so one would expect inverse LETFs to perform poorly even independent of all the headwinds discussed in Sections~\ref{sec:volatility_drag} and~\ref{sec:fees_financing}.

\begin{figure}[!htbp]\centering
\includegraphics[
height=3.0in,
alt={Two line charts show cumulative investor flows into leveraged ETF categories. Over the full sample, leveraged equity-index ETFs receive the largest cumulative inflows. In the post-2023 panel, single-stock LETFs receive large positive inflows, while leveraged equity-index ETFs experience substantial outflows.}
]{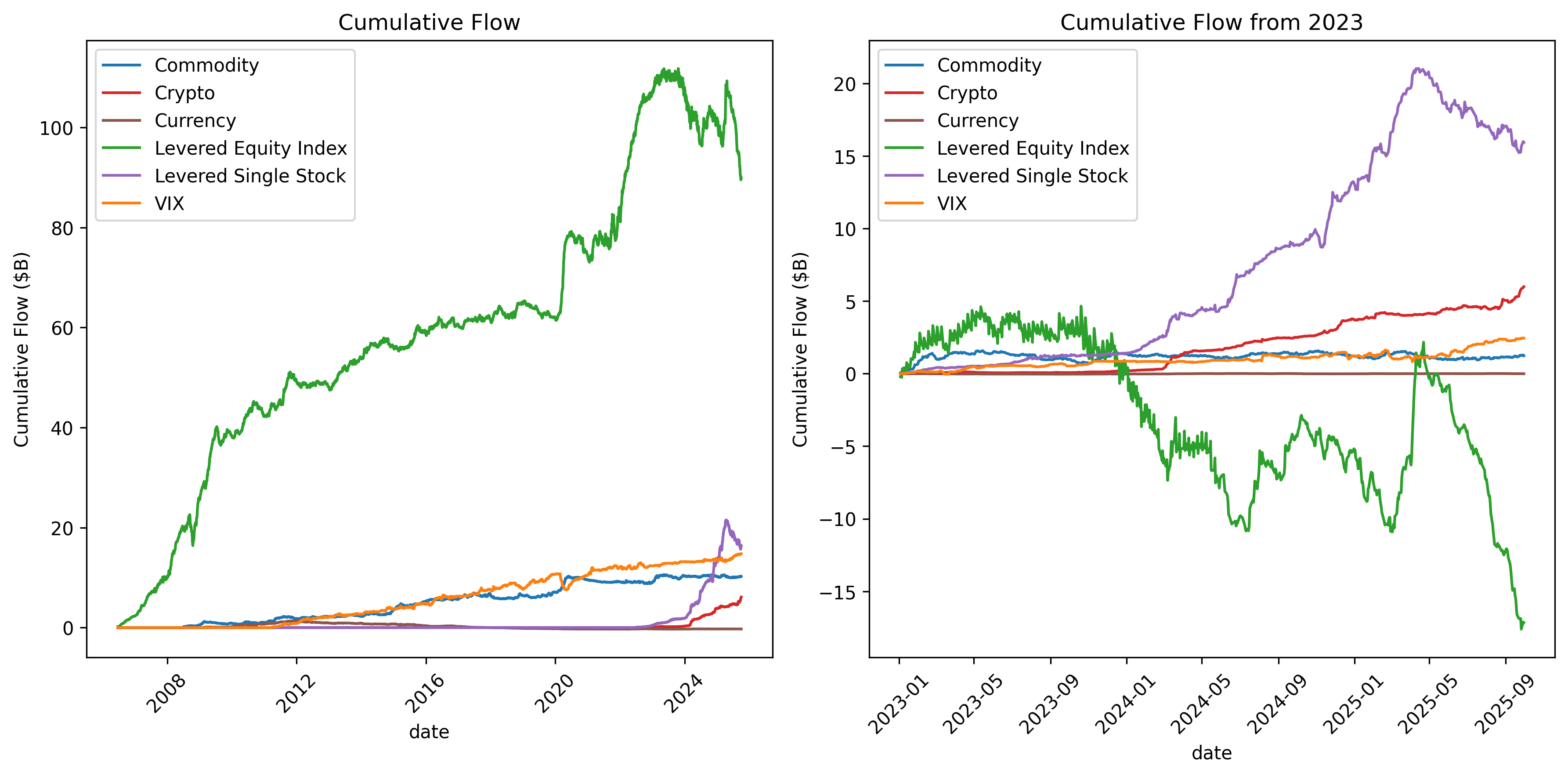}
\caption[]{\textbf{Flows by Leveraged ETF Type.} The figure reports cumulative investor flows by leveraged ETF type. The left panel covers the full sample period, while the right panel begins in 2023. Daily flows are defined in equation~\ref{eq:flow}.}
\figalt{Two line charts show cumulative investor flows into leveraged ETF categories. Over the full sample, leveraged equity-index ETFs receive the largest cumulative inflows. In the post-2023 panel, single-stock LETFs receive large positive inflows, while leveraged equity-index ETFs experience substantial outflows.}
\label{fig:flow_by_etf_type}
\end{figure}

The realized gains in long LETFs came mainly from equity products. Long leveraged equity-index ETFs generated \$117.8 billion in investor PnL, while long leveraged single-stock ETFs generated \$13.3 billion. These gains qualify the common view that LETFs are generally unsuitable for longer-horizon investors \citep{cheng2009dynamics}: at least for broad equity-index LETFs over our sample, realized equity returns were large enough to overcome volatility drag, fees, and financing costs. Commodity and VIX products generated losses. Despite the large gains in equity-index products, equity-index LETFs have negative net flows, indicating that investors redeemed shares after strong performance. Figure~\ref{fig:flow_by_etf_type} shows the same pattern: long equity-index LETFs experienced sustained outflows over the 2023--2025 period, while long single-stock LETFs continued to attract inflows.

These aggregate results show that broad equity-index LETFs generated large dollar gains for investors despite volatility drag, fees, and financing costs. But aggregate PnL does not tell us whether the LETF wrapper added value relative to simply holding the asset being levered. To answer that question, we need a counterfactual, which we establish in the next subsection.

\subsection{Returns Relative to the Tracked Asset}
\label{subsec:counterfactual_underlying}

We compare long LETF investor outcomes to a simple counterfactual: investing the same dollars, at the same times, in the unlevered asset the LETF tracks. This asks whether investors would have done better or worse with ordinary unlevered exposure to the same asset, holding fixed the timing of their flows. We restrict the analysis to long LETFs for which we can identify a daily return series for the underlying asset or a close unlevered ETF proxy.\footnote{For long products, the natural unlevered benchmark is usually observable. For example, an S\&P 500 LETF can be compared to the return on the S\&P 500 index or an unlevered S\&P 500 ETF. For inverse products, there is no analogous public benchmark that gives the realized return to shorting the underlying asset without using the inverse LETF itself. Constructing such a benchmark would require assumptions about shorting costs, margin requirements, collateral, and financing, so we exclude inverse products from this counterfactual analysis.}

Other natural benchmarks are possible, including levered strategies implemented with margin, futures, options, or less frequent rebalancing. Our goal is not to compare all ways of obtaining leverage, but to ask whether the LETF wrapper added value relative to ordinary unlevered exposure to the same asset.

The counterfactual, like our baseline Investor PnL measure, does not incorporate investor-level taxes. When investors put money into the LETF, our counterfactual exercise puts the same dollar amount into the unlevered asset. When investors take money out, we reduce the counterfactual position pro rata at the same redemption rate. For example, if investors redeem 10\% of the LETF's assets on a given day, we redeem 10\% of the counterfactual position as well. Using percentages for outflows keeps the comparison well-defined even when the LETF and the unlevered asset have earned different returns.\footnote{See Appendix Section~\ref{app:counterfactual_details} for the exact accounting details of our counterfactual exercise.}

Panel B of Table~\ref{tab:investor_pnl_combined} reports the results. Leveraged equity-index ETFs generated \$107.6 billion in investor PnL in the matched sample, compared with \$61.1 billion in the underlying-asset counterfactual. Leveraged single-stock ETFs also generated positive investor PnL, but the incremental gain relative to the underlying stocks is much smaller: \$13.1 billion in LETF investor PnL compared with \$11.7 billion in the underlying counterfactual. Commodity and VIX LETFs underperformed their underlying counterfactuals.

\begin{figure}[htb!]
\centering
\includegraphics[
width=0.75\linewidth,
alt={Scatter plot of two-times long LETFs showing the annualized log-return gap relative to twice the log return of the underlying asset against underlying asset daily volatility. The points slope downward: higher underlying volatility is associated with more negative realized log-return gaps. A red frictionless volatility-drag prediction line also slopes downward and becomes more negative as volatility rises.}
]{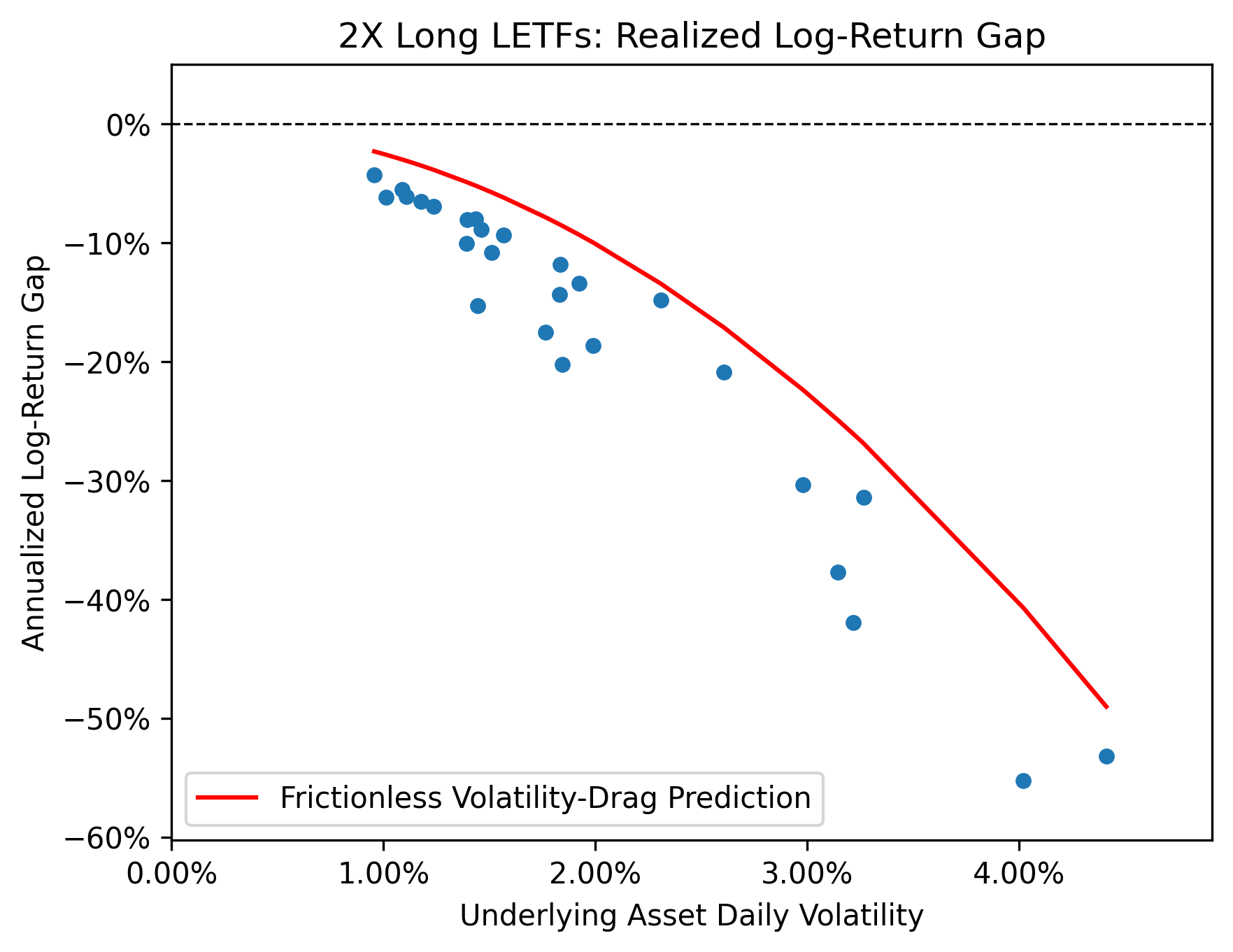}
\caption{\textbf{Realized Log-Return Gap for $2\times$ Long LETFs.} The figure plots each $2\times$ long LETF's annualized log-return gap relative to twice the log return of its underlying asset against the underlying asset's realized daily volatility. The red line shows the frictionless volatility-drag prediction from Equation~\ref{eq:lev_drag_ret}.}
\figalt{Scatter plot of $2\times$ long LETFs showing the annualized log-return gap relative to twice the log return of the underlying asset against underlying asset daily volatility. The points slope downward: higher underlying volatility is associated with more negative realized log-return gaps. A red frictionless volatility-drag prediction line also slopes downward and becomes more negative as volatility rises.}
\label{fig:realized_drag_2x}
\end{figure}

The bottom section of Panel B groups the same observations by the daily volatility of the underlying asset. In the lowest-volatility bucket, where underlying daily volatility is below 1.5\%, LETFs generated \$78.9 billion in investor PnL, compared with \$39.9 billion in the underlying counterfactual. In the 1.5\% to 2\% bucket, LETFs again outperformed the underlying counterfactual. At higher volatility levels, the sign flips. For assets with daily volatility between 2\% and 3\%, LETFs generated negative investor PnL even though the underlying counterfactual generated positive gains. For assets with daily volatility above 3\%, LETF investor PnL was \$6.4 billion, compared with \$8.3 billion in the underlying counterfactual. This relationship between volatility and investor PnL is consistent with the fund-level evidence in Figure~\ref{fig:realized_drag_2x}, which shows that $2\times$ long LETF log-return gaps become more negative as underlying volatility rises and broadly follow the frictionless volatility-drag prediction from Equation~\ref{eq:lev_drag_ret}.

\begin{figure}[t]
\centering
\includegraphics[
height=3in,
alt={Two line charts show cumulative investor outperformance of long LETFs relative to investments in the underlying assets, grouped by underlying daily volatility. LETFs tied to lower-volatility assets generate positive cumulative excess PnL, while products tied to higher-volatility assets underperform. The normalized panel shows the same pattern after scaling by the number of observations in each volatility bucket.}
]{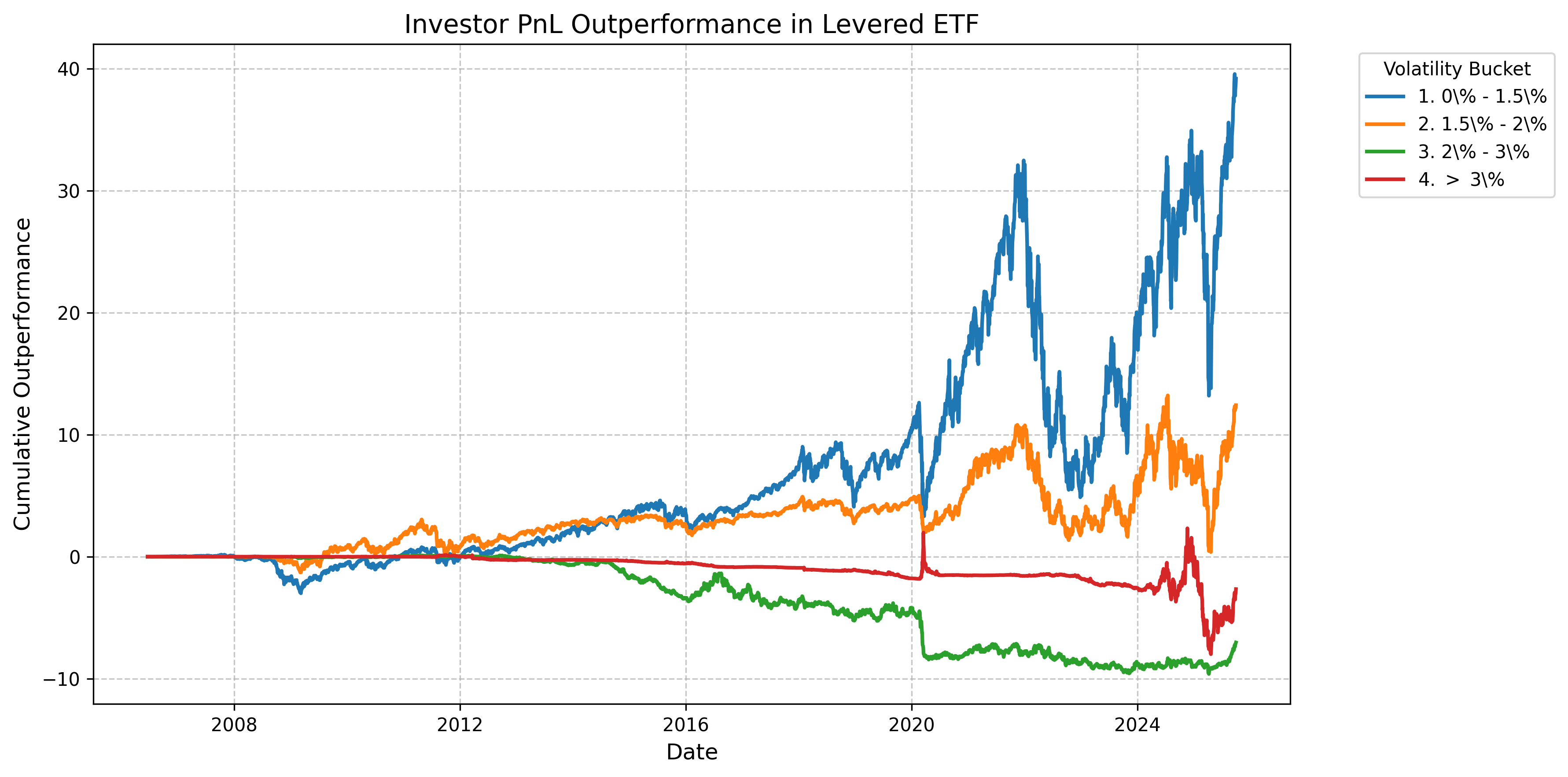}
\includegraphics[height=3in]{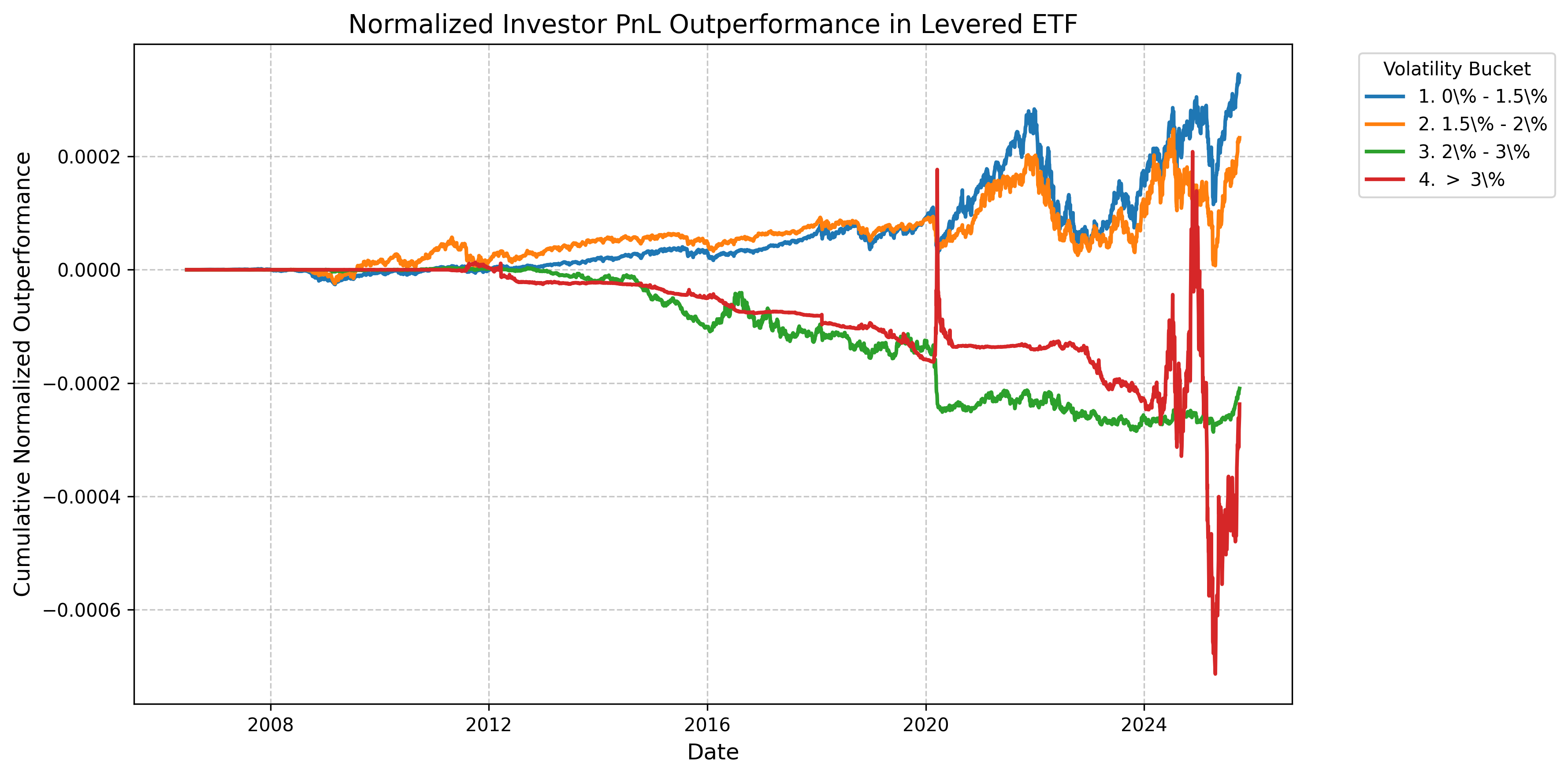}
\caption{\textbf{Investor Excess PnL by Underlying Volatility.} The figure reports cumulative investor excess PnL in long LETFs with available underlying return data, grouped by the daily volatility of the underlying asset. The first panel displays raw outperformance relative to the underlying counterfactual, while the second normalizes by the number of observations in each bucket.}
\figalt{Two line charts show cumulative investor outperformance of long LETFs relative to investments in the underlying assets, grouped by underlying daily volatility. LETFs tied to lower-volatility assets generate positive cumulative excess PnL, while products tied to higher-volatility assets underperform. The normalized panel shows the same pattern after scaling by the number of observations in each volatility bucket.}
\label{fig:letf_outperf_by_vol}
\end{figure}

Figure~\ref{fig:letf_outperf_by_vol} shows the same comparison over time. The excess investor PnL of LETFs over the underlying counterfactual is concentrated in lower-volatility assets. Higher-volatility LETFs do not show the same persistent advantage. This lines up with the frictionless-market breakeven-return calculations in Section~\ref{sec:volatility_drag}: as the underlying asset becomes more volatile, the LETF needs a much larger underlying return to overcome volatility drag, fees, and financing costs. Broad equity-index LETFs performed well because broad equity returns were strong over our sample period, volatility was relatively low, and financing costs were lower for these products. Single-stock LETFs and other products tied to more volatile assets face a higher breakeven hurdle for the reasons discussed in Section~\ref{sec:volatility_drag}: higher volatility and higher financing costs imply a much higher breakeven return.

\subsection{Return Decomposition}
\label{subsec:investor_return_decomposition}

Aggregate investor PnL is measured in dollars, so it will be dominated by high-AUM funds and by time periods when capital invested in those funds is large. Ultimately, however, we want to know whether investor returns come from the products themselves, from investors choosing better products, or from the timing of flows within products. We therefore decompose investor performance into three pieces: product returns, fund selection, and within-fund timing.

To separate product performance from investor timing, we compare two returns for each fund. The first is the fund's time-weighted return, which treats each day equally and does not depend on when investors add or remove capital. The second is an investor-capital-weighted return, which puts more weight on days when more investor capital is in the fund. The difference between the two measures captures within-fund timing: whether investors tend to have more money in the fund before good days or bad days.

We then aggregate these components within each ETF type. The raw fund return is the equal-weighted average return across funds. The fund selection effect captures whether investor capital is allocated to better-performing funds within a category. The investor timing effect captures whether investor capital within a fund is present before higher- or lower-return days. Thus, at the group level,
\[
R^{Investor}
=
R^{Raw}
+
\text{Selection Effect}
+
\text{Timing Effect}.
\]
Appendix Section~\ref{app:return_decomp_details} describes the construction of investor capital and the exact return formulas.

Panel A of Table~\ref{tab:investor_return_decomp_combined} reports the decomposition for long LETFs with at least two years of trading history. Leveraged equity-index LETFs have an investor return of 7.8\%, consisting of a 4.0\% raw fund return, a positive 12.6\% selection effect, and a negative 8.8\% timing effect. Investors allocate capital to better-performing equity-index LETFs, but their timing within those funds reduces returns.

Single-stock LETFs were first launched shortly before a period of unusually high market returns, including especially strong returns for many of their underlying stocks. This means that our decomposition is best interpreted \textit{within} product categories rather than as a direct comparison of average returns across single-stock and equity-index LETFs. Leveraged single-stock LETFs have a raw fund return of 46.5\% and a positive selection effect of 43.1\%, so investor capital is concentrated in higher-return products within the category. But the investor timing effect is $-48.3\%$. In other words, investors allocated capital to single-stock LETFs that performed well on average, but they tended to have especially poor timing within those funds. 

Panel B of Table~\ref{tab:investor_return_decomp_combined} repeats the decomposition for unlevered ETFs. For regular ETFs the selection and timing components are much smaller. Specifically, we find that for broad-based equity ETFs, the investor return is 13.3\%, the raw fund return is 10.8\%, the selection effect is 0.6\%, and the timing effect is 1.9\%. We find similar patterns for global and sub-index equity ETFs. Investor returns in unlevered ETFs, therefore, are much closer to the unconditional returns of the products themselves.

Overall, we believe this decomposition reinforces the distinction between the products that worked best in our sample and the products that have been growing rapidly over the last few years. Broad equity-index LETFs benefited from diversified equity exposure, relatively low volatility, and lower financing costs.  Further, end-investor timing reduced returns, but the effect was small. Single-stock LETFs, however, looked different: investors selected products with high average returns, but within-fund timing was strongly negative, and the same volatility that attracts attention, AUM, and trading volume also raises breakeven returns through volatility drag, fees, and financing costs.

\section{Conclusion}
\label{sec:conclusion}

LETFs make leverage easy to access. Investors can obtain levered exposure in an ETF wrapper without opening a margin account, trading derivatives, or rebalancing the position themselves. And, we show that this access is not inherently bad: in our sample, it was valuable mainly for broad equity-index products, which generated over \$100 billion in investor PnL, compared with \$61.1 billion from investing the same dollars, at the same times, in the corresponding unlevered assets. When broad equity returns are high, volatility is relatively low, and financing costs are modest, the benefits of leverage can outweigh volatility drag, fees, and financing costs.

But whether easy access to leverage benefits investors depends on what asset is actually being levered. A $2\times$ product on a broad equity index and a $2\times$ product on a volatile single stock both target twice the benchmark's \textit{daily} return, but they are not close substitutes for investors with longer holding periods. When the same leverage multiple is applied to a more volatile asset, the breakeven return becomes much harder to reach, as volatility drag rises with realized variance. On top of this, financing costs also tend to be higher for more volatile underlying assets. The relevant question is therefore not whether investors should have easy access to leverage in the ETF wrapper, but the interaction between leverage and the asset being levered.

This distinction matters because product entry is moving the LETF market toward the types of products where breakeven returns are likely to be the highest. Single-stock LETFs in our sample are concentrated in volatile recent winners, and products tied to more volatile stocks attract more AUM and trade more heavily after launch. This is not surprising. Volatile winners are exactly the kind of assets investors notice, talk about, and want levered exposure to.  Moreover, LETFs on these underlying assets may cater directly to retail investors' biases e.g., demand for lottery-like payoffs.   But these are also precisely the assets where leverage creates the largest structural headwinds for longer-horizon returns.

While we show that investors had large realized gains in broad equity-index LETFs, we argue that it is not driven by LETF investors' market-timing skill.  On average, LETF investors buy after recent losses and sell after recent gains, but these flows do not predict subsequent returns. In our return decomposition, we show that investors allocate capital to better-performing LETFs \textit{across} funds, but their timing \textit{within} funds is poor. Thus, investor timing does not appear to be the force that makes LETFs work when they work, nor does it appear to offset the larger breakeven return hurdles in more volatile products.

From a regulatory perspective, disclosing an LETF's \textit{daily}-return objective tells investors how the product operates, but not what they are likely to earn over their expected holding period. Those returns depend on the asset being levered, its volatility, the product's fees and financing costs, and how the investor trades it. When evaluating LETFs, therefore, the concern is not whether easy access to leverage is good or bad, but which assets are being levered. As LETFs move from broad indexes toward single-stock and other targeted exposures, retail-accessible leverage is increasingly being applied to assets with strong investor demand but also high breakeven returns. In other words, financial innovation has made these exposures easy to buy, but their longer-horizon economics are still not easy to evaluate.

\clearpage
\section*{Figures and Tables}
\addcontentsline{toc}{section}{Figures and Tables}
\suppressfloats[t] 
\FloatBarrier

\begin{table}[!htb]
\begin{center}

\begin{tabular}{P{5.1cm}|P{3.8cm}P{3.8cm}}
\hline\hline
\multicolumn{3}{c}{\textbf{Panel A: Breakeven Returns}} \\
\hline
Underlying Volatility
& \multicolumn{2}{c}{Breakeven Return} \\
\cline{2-3}
& L=2 & L=3 \\
\hline
$1.0\%$ & $8.4\%$ & $8.9\%$ \\
$2.0\%$ & $16.9\%$ & $22.0\%$ \\
$3.0\%$ & $32.6\%$ & $47.4\%$ \\
$4.0\%$ & $58.1\%$ & $92.0\%$ \\
$5.0\%$ & $98.4\%$ & $169.8\%$ \\
$6.0\%$ & $161.7\%$ & $308.9\%$ \\
$7.0\%$ & $263.2\%$ & $568.4\%$ \\
\hline
\end{tabular}

\vspace{0.35cm}

\begin{tabular}{P{3.0cm}|P{2.2cm}P{2.2cm}|P{2.2cm}P{2.2cm}}
\hline
\multicolumn{5}{c}{\textbf{Panel B: Underlying Volatility and Volatility Drag}} \\
\hline
Year
& \multicolumn{2}{c|}{Average Volatility}
& \multicolumn{2}{c}{Annual Volatility Drag} \\
\cline{2-5}
& Levered Equity Index
& Levered Single Stock
& Levered Equity Index
& Levered Single Stock \\
\hline
2022 & $1.87\%$ & $3.49\%$ & $-8.43\%$ & $-26.41\%$ \\
2023 & $1.17\%$ & $2.91\%$ & $-3.41\%$ & $-19.20\%$ \\
2024 & $1.20\%$ & $3.84\%$ & $-3.55\%$ & $-31.07\%$ \\
2025 & $1.35\%$ & $3.71\%$ & $-4.52\%$ & $-29.31\%$ \\
\hline\hline
\end{tabular}

\end{center}

\caption[]{\textbf{Breakeven Returns, Underlying Volatility, and Volatility Drag.} Panel A shows the annual underlying return required for a levered ETF of a given leverage ratio ($L$) to have a return equal to the underlying asset. If the underlying asset has a higher return than this breakeven return during the year, the levered ETF will outperform the underlying asset. If the underlying return is below this breakeven return, the levered ETF will underperform it. We assume a risk-free rate of 4.0\%, an expense ratio of 1.0\%, and a swap financing spread of 0.5\%, all annualized. Panel B reports the 21-day realized daily volatility of assets tracked by leveraged ETFs, weighted by AUM.
The final two columns report the corresponding annual return effect from volatility drag for a $2\times$ long LETF.}
\label{tab:breakeven_vol_drag}
\end{table}
\FloatBarrier

\begin{table}
\begin{center}

\begin{tabular}{P{6.75cm}|P{5.25cm}}
\hline\hline
\multicolumn{2}{c}{\textbf{Panel A: Expense Ratios}} \\
\hline
ETF Type & Expense Ratio\\
\hline
Bonds & 0.63\% \\
Commodity & 1.03\% \\
Crypto & 1.57\% \\
Currency & 0.95\% \\
Levered Equity Index & 0.90\% \\
Levered Single Stock & 1.03\% \\
VIX & 1.60\% \\
\hline
\end{tabular}

\vspace{0.35cm}

\begin{tabular}{P{4.5cm}|P{2.2cm}P{2.2cm}P{2.2cm}}
\hline
\multicolumn{4}{c}{\textbf{Panel B: Funding Spreads}} \\
\hline
 & \multicolumn{3}{c}{\textbf{Market-Cap Weighted}} \\
\hline
Type & Levered Single Stock & Levered Equity Index & Levered Equity Index \\
Leverage & $2.0$ & $2.0$ & $3.0$ \\
Annualized Funding Spread & $9.8\%$ & $0.5\%$ & $0.9\%$ \\
 & $(9.37)$ & $(2.48)$ & $(4.65)$ \\
\hline
Observations & $4,669$ & $34,368$ & $53,727$ \\
\hline
\multicolumn{4}{c}{} \\
[-1.8ex]
\hline
 & \multicolumn{3}{c}{\textbf{Equal Weighted}} \\
\hline
Type & Levered Single Stock & Levered Equity Index & Levered Equity Index \\
Leverage & $2.0$ & $2.0$ & $3.0$ \\
Annualized Funding Spread & $6.8\%$ & $0.4\%$ & $0.5\%$ \\
 & $(9.16)$ & $(1.50)$ & $(4.78)$ \\
\hline
Observations & $4,669$ & $34,368$ & $53,727$ \\
\hline\hline
\end{tabular}

\end{center}

\caption[]{\textbf{Expense Ratios and Funding Spreads.} Panel A reports AUM-weighted expense ratios for LETFs in our universe. Panel B reports annualized funding-spread estimates for leveraged single-stock and equity-index ETFs under market-cap and equal-weighting schemes. The ideal daily return before the funding spread is $R_{L,\text{ideal}} = L R_u - (L - 1) R_{rf} - \text{DailyExpenseRatio}$. We estimate $R_L - R_{L,\text{ideal}} = -\beta_1 (L-1)/252 + \epsilon$, so that $\beta_1$ is our estimated annualized funding spread. The sample is restricted to ETFs with daily tracking error relative to $L R_u$ below 50 basis points, excludes the top 2\% of observations with the largest absolute deviations from the ideal return, and excludes LETFs whose leverage ratio changed over the product life.}
\label{tab:costs}
\end{table}
\FloatBarrier

\FloatBarrier

\begin{table}[t]
\centering

\resizebox{0.95\linewidth}{!}{
\begin{tabular}{P{5.5cm}|P{2.2cm}P{2.2cm}P{2.2cm}}
\hline\hline
ETF Type & 2000s & 2010s & 2020s\\
\hline
\multicolumn{4}{c}{\textbf{Panel A: ETF Count}} \\
\hline
Commodity & 5 & 10 & 9 \\
Crypto & - & - & 10 \\
Currency & 1 & 1 & 1 \\
Levered Equity Index & 39 & 85 & 92 \\
Levered Single Stock & - & - & 71 \\
Unlevered & 40 & 54 & 57 \\
VIX & - & 2 & 3 \\
\hline
\multicolumn{4}{c}{\textbf{Panel B: Total Daily Turnover}} \\
\hline
Commodity & 10.78\% & 6.79\% & 8.57\% \\
Crypto & - & - & 20.86\% \\
Currency & 6.69\% & 3.19\% & 4.37\% \\
Levered Equity Index & 71.59\% & 20.35\% & 16.37\% \\
Levered Single Stock & - & - & 35.08\% \\
Unlevered Bottom Volume & 1.87\% & 0.91\% & 0.75\% \\
Unlevered Top Volume & 22.18\% & 8.14\% & 5.51\% \\
VIX & - & 56.39\% & 30.06\% \\
\hline
\multicolumn{4}{c}{\textbf{Panel C: Mean Turnover}} \\
\hline
Commodity & 14.45\% & 9.67\% & 16.06\% \\
Crypto & - & - & 20.07\% \\
Currency & 6.69\% & 3.19\% & 4.37\% \\
Levered Equity Index & 37.54\% & 10.53\% & 8.34\% \\
Levered Single Stock & - & - & 44.26\% \\
Unlevered Bottom Volume & 1.93\% & 0.93\% & 0.93\% \\
Unlevered Top Volume & 22.01\% & 8.36\% & 5.83\% \\
VIX & - & 106.26\% & 75.61\% \\
\hline\hline
\end{tabular}
}

\caption[]{\textbf{ETF Counts and Turnover by Product Type.} Panel A reports the number of ETFs in each category. Panel B reports total daily turnover, defined as aggregate dollar volume divided by aggregate AUM within each category. Panel C reports mean product-level daily turnover. Unlevered ETFs are split into bottom-volume and top-volume groups within our universe of unlevered ETFs.}
\label{tab:turnover_by_period_type}

\end{table}
\FloatBarrier

\begin{table}
\centering

\begin{tabular}{P{4.2cm}|P{1.8cm}P{1.9cm}P{1.9cm}P{2.0cm}P{2.0cm}}
\hline\hline
\multicolumn{6}{c}{\textbf{Panel A: Launch Date Characteristics}} \\
\hline
Rank Column & ETF Launches & Mean Percentile & Median Percentile & Pct. Above 75th Percentile & Pct. Above 90th Percentile\\
\hline
Underlying Volatility & $43$ & $86.3$ & $93.8$ & $86.0$ & $65.1$ \\
Underlying Past Return & $40$ & $70.4$ & $91.7$ & $62.5$ & $52.5$ \\
\hline\hline
\end{tabular}

\vspace{0.35cm}

\begin{tabular}{P{8.8cm}|P{3.1cm}P{3.1cm}}
\hline\hline
\multicolumn{3}{c}{\textbf{Panel B: Average AUM Regressions}} \\
\hline
 & (1) & (2)\\
\hline
Intercept & $8.60$ & $7.44$ \\
 & $(2.98)$ & $(1.23)$ \\
Underlying Daily Volatility & $51.78$ & $76.80$ \\
 & $(4.02)$ & $(3.84)$ \\
Log Underlying Market Cap & $0.11$ & $0.30$ \\
 & $(1.05)$ & $(1.46)$ \\
Log Number of Fund Days & $0.80$ &  \\
 & $(4.26)$ &  \\
\hline
Observations & $68$ & $40$ \\
Adjusted $R^2$ & $0.22$ & $0.21$ \\
\hline\hline
\end{tabular}

\caption[]{\textbf{Single-Stock LETF Launch Characteristics and AUM.} Panel A reports launch-date underlying-stock characteristics for single-stock LETFs launched in February 2023 or later, restricting the sample to products whose underlying asset is among the top 1,000 stocks in CRSP by market capitalization on the launch date. Market capitalization is measured using a 21-day average. Launch-date volatility is measured as the rolling 21-day standard deviation of daily returns, and past returns are measured as cumulative underlying-stock returns over the 126 trading days ending on the day before launch. For each launch date, we rank the underlying stock's volatility and past 126-day return relative to this top-1,000-stock universe. Panel B reports fund-level OLS regressions of log average AUM on underlying-stock characteristics for long leveraged single-stock ETFs, with heteroskedasticity-robust t-statistics in parentheses. The dependent variable is the log of average daily AUM over the fund's sample period. Underlying Daily Volatility is the underlying stock's average daily volatility over the fund's sample period, measured in decimal units, and Log Underlying Market Cap is the log of the underlying stock's average market capitalization. Column (1) includes all funds with non-missing underlying asset data and controls for fund age using the log number of fund trading days. Column (2) restricts the sample to funds with at least 252 trading days.}
\label{tab:single_stock_launch_aum}

\end{table}
\FloatBarrier

\FloatBarrier

\begin{table}[t]
\centering
\scriptsize
\setlength{\tabcolsep}{3.5pt}

\resizebox{0.98\linewidth}{!}{
\begin{tabular}{P{4.4cm}|P{2.0cm}P{2.0cm}P{2.0cm}P{2.0cm}}
\hline\hline
\multicolumn{5}{c}{\textbf{Panel A: Long LETFs}} \\
\hline
 & (1) & (2) & (3) & (4)\\
\hline
Lagged 1-Day Return & $-4.02$ & $-4.81$ & $-4.72$ & $-2.16$ \\
 & $(-14.88)$ & $(-12.42)$ & $(-21.99)$ & $(-5.89)$ \\
Lagged 5-Day Return & $-2.59$ & $-3.34$ & $-2.70$ & $-1.37$ \\
 & $(-20.02)$ & $(-18.07)$ & $(-25.45)$ & $(-8.84)$ \\
Lagged 21-Day Return & $-0.46$ & $-0.61$ & $-0.33$ & $-0.31$ \\
 & $(-7.25)$ & $(-6.24)$ & $(-6.34)$ & $(-3.92)$ \\
Lagged 63-Day Return & $-0.07$ & $-0.02$ & $-0.04$ & $-0.10$ \\
 & $(-1.69)$ & $(-0.28)$ & $(-1.40)$ & $(-1.92)$ \\
Lagged 126-Day Return & $0.07$ & $-0.05$ & $0.09$ & $0.18$ \\
 & $(3.06)$ & $(-1.37)$ & $(4.91)$ & $(4.78)$ \\
Intercept & $-0.02$ & $-0.04$ & $-0.00$ & $-0.01$ \\
 & $(-1.86)$ & $(-3.83)$ & $(-0.33)$ & $(-0.78)$ \\
\hline
Observations & $323,786$ & $30,119$ & $161,295$ & $38,758$ \\
Adjusted $R^2$ & $0.07$ & $0.17$ & $0.09$ & $0.05$ \\
Period & Entire Sample & 2006 to 2011 & 2012 to 2020 & 2024 and 2025 \\
\hline
\multicolumn{5}{c}{\textbf{Panel B: Unlevered ETFs}} \\
\hline
 & (1) & (2) & (3) & (4)\\
\hline
Lagged 1-Day Return & $2.23$ & $2.84$ & $5.29$ & $0.02$ \\
 & $(3.24)$ & $(4.23)$ & $(6.93)$ & $(0.02)$ \\
Lagged 5-Day Return & $2.42$ & $1.85$ & $2.03$ & $2.53$ \\
 & $(7.05)$ & $(5.14)$ & $(5.33)$ & $(4.14)$ \\
Lagged 21-Day Return & $-0.06$ & $0.73$ & $0.36$ & $-0.50$ \\
 & $(-0.33)$ & $(3.78)$ & $(1.91)$ & $(-1.64)$ \\
Lagged 63-Day Return & $0.34$ & $-0.53$ & $0.05$ & $0.37$ \\
 & $(2.90)$ & $(-3.86)$ & $(0.43)$ & $(2.02)$ \\
Lagged 126-Day Return & $0.43$ & $0.03$ & $0.75$ & $0.38$ \\
 & $(5.82)$ & $(0.39)$ & $(8.97)$ & $(2.35)$ \\
Intercept & $0.06$ & $0.00$ & $-0.02$ & $-0.05$ \\
 & $(10.34)$ & $(0.41)$ & $(-2.83)$ & $(-3.19)$ \\
\hline
Observations & $229,404$ & $42,575$ & $118,984$ & $23,156$ \\
Adjusted $R^2$ & $0.01$ & $0.01$ & $0.01$ & $0.00$ \\
Period & Entire Sample & 2006 to 2011 & 2012 to 2020 & 2024 and 2025 \\
\hline\hline
\end{tabular}
}

\caption{\textbf{Regression of ETF Flows on Lagged Returns.} The table reports regressions of normalized daily flows on lagged fund returns over multiple horizons. Panel A reports results for long LETFs, and Panel B reports results for unlevered ETFs. Daily flows are winsorized at the 2.5\% and 97.5\% levels within fund and then standardized within fund. Lagged returns are winsorized at the same thresholds. All returns are measured using information available before the flow date. Regressions are estimated separately by sample period, with standard errors clustered by date. Observations are weighted by previous-day AUM.}
\label{tab:flow_timing_regressions}
\end{table}
\FloatBarrier

\begin{table}[t]
\begin{center}
\resizebox{0.90\linewidth}{!}{
\begin{tabular}{P{5.0cm}|P{2.2cm}P{2.2cm}P{2.2cm}P{2.2cm}}
\hline\hline
 & (1) & (2) & (3) & (4)\\
\hline
Lagged 1-Day Return & $-4.02$ & $-4.85$ & $-4.90$ & $-2.04$ \\
 & $(-14.38)$ & $(-12.38)$ & $(-22.52)$ & $(-4.65)$ \\
Lagged 5-Day Return & $-2.49$ & $-3.38$ & $-2.81$ & $-1.05$ \\
 & $(-18.62)$ & $(-18.14)$ & $(-26.74)$ & $(-5.65)$ \\
Lagged 21-Day Return & $-0.47$ & $-0.59$ & $-0.30$ & $-0.32$ \\
 & $(-7.41)$ & $(-6.03)$ & $(-5.95)$ & $(-3.19)$ \\
Lagged 63-Day Return & $0.06$ & $0.11$ & $-0.04$ & $0.06$ \\
 & $(1.55)$ & $(1.55)$ & $(-1.33)$ & $(0.77)$ \\
Lagged 126-Day Return & $0.01$ & $-0.14$ & $0.09$ & $0.10$ \\
 & $(0.36)$ & $(-3.36)$ & $(4.78)$ & $(1.99)$ \\
Future 1-Day Return & $-0.11$ & $-0.45$ & $-0.29$ & $0.11$ \\
 & $(-0.41)$ & $(-1.17)$ & $(-1.33)$ & $(0.28)$ \\
Future 5-Day Return & $-0.11$ & $-0.27$ & $-0.09$ & $0.05$ \\
 & $(-0.77)$ & $(-1.43)$ & $(-0.81)$ & $(0.28)$ \\
Future 21-Day Return & $-0.12$ & $-0.20$ & $-0.14$ & $-0.21$ \\
 & $(-1.62)$ & $(-2.00)$ & $(-2.72)$ & $(-2.03)$ \\
Future 63-Day Return & $0.04$ & $-0.05$ & $-0.04$ & $0.16$ \\
 & $(0.91)$ & $(-0.72)$ & $(-1.46)$ & $(2.20)$ \\
Future 126-Day Return & $-0.07$ & $-0.20$ & $0.05$ & $-0.02$ \\
 & $(-2.47)$ & $(-4.23)$ & $(3.28)$ & $(-0.38)$ \\
Intercept & $0.01$ & $-0.04$ & $-0.00$ & $0.02$ \\
 & $(1.58)$ & $(-4.25)$ & $(-0.06)$ & $(1.05)$ \\
\hline
Observations & $307,910$ & $30,119$ & $159,909$ & $27,040$ \\
Adjusted $R^2$ & $0.07$ & $0.18$ & $0.09$ & $0.03$ \\
Period & Entire Sample & 2006 to 2011 & 2012 to 2020 & 2024 and 2025 \\
\hline\hline
\end{tabular}
}
\end{center}
\caption{\textbf{Regression of Long LETF Flows on Lagged and Future Returns.} For long LETFs, we regress normalized daily flows on lagged and subsequent fund returns over multiple horizons. Daily flows are winsorized at the 2.5\% and 97.5\% levels within fund and then standardized within fund. Return variables are winsorized at the same thresholds. Lagged returns are measured using information available before the flow date, while future returns are measured after the flow date. Regressions are estimated separately by sample period, with standard errors clustered by date to account for cross-sectional correlation across ETFs on the same day. Observations are weighted by previous-day AUM.}
\label{tab:flowTimingRegressionFutureletf1}
\end{table}
\FloatBarrier

\begin{table}[t]
\begin{center}
\resizebox{0.98\linewidth}{!}{
\begin{tabular}{P{4.2cm}|P{1.8cm}P{2.2cm}P{2.5cm}P{2.0cm}P{1.8cm}}
\hline\hline
Group & ETF Count & Investor PnL(\$B) & Underlying Investor PnL(\$B) & Net Flow(\$B) & AUM(\$B)\\
\hline
\multicolumn{6}{c}{\textbf{Panel A: Aggregate LETF Market}} \\
\hline
\multicolumn{6}{c}{\textit{By leverage type}} \\
\hline
Long & 193 & 114.8 & -- & 22.8 & 137.6 \\
Inverse & 112 & -80.3 & -- & 93.4 & 13.1 \\
\hline
\multicolumn{6}{c}{\textit{Long LETFs by ETF type}} \\
\hline
Commodity & 10 & -5.7 & -- & 8.8 & 3.0 \\
Crypto & 10 & 2.1 & -- & 5.1 & 7.1 \\
Currency & 1 & -0.0 & -- & 0.1 & 0.1 \\
Levered Equity Index & 98 & 117.8 & -- & -17.3 & 100.4 \\
Levered Single Stock & 71 & 13.3 & -- & 12.5 & 25.7 \\
VIX & 3 & -12.6 & -- & 13.8 & 1.2 \\
\hline
\multicolumn{6}{c}{\textbf{Panel B: Long LETFs with Underlying Returns}} \\
\hline
\multicolumn{6}{c}{\textit{By ETF type}} \\
\hline
Commodity & $8$ & $-6.1$ & $-2.1$ & $8.9$ & $2.8$ \\
Levered Equity Index & $47$ & $107.6$ & $61.1$ & $-14.1$ & $93.5$ \\
Levered Single Stock & $53$ & $13.1$ & $11.7$ & $11.4$ & $24.5$ \\
VIX & $1$ & $-3.0$ & $-1.6$ & $3.0$ & -- \\
\hline
\multicolumn{6}{c}{\textit{By underlying daily volatility bucket}} \\
\hline
1. 0\% - 1.5\% & $30$ & $78.9$ & $39.9$ & $-9.5$ & $69.4$ \\
2. 1.5\% - 2\% & $19$ & $30.1$ & $17.6$ & $-8.6$ & $21.5$ \\
3. 2\% - 3\% & $22$ & $-3.8$ & $3.2$ & $11.0$ & $7.1$ \\
4. $>$ 3\% & $38$ & $6.4$ & $8.3$ & $16.4$ & $22.8$ \\
\hline\hline
\end{tabular}
}
\end{center}
\caption{\textbf{Investor PnL and Performance Relative to the Underlying Asset.} The table reports investor PnL, net flows, and assets under management for LETFs in our sample. Panel A reports aggregate results for all LETFs by leverage type and for long LETFs by ETF type. Panel B restricts the sample to long LETFs with available underlying asset returns and adds the counterfactual investor PnL from investing the same flows directly in the underlying asset. Within Panel B, results are reported by ETF type and by underlying daily volatility bucket. Dollar values are reported in billions.}
\label{tab:investor_pnl_combined}
\end{table}
\FloatBarrier

\FloatBarrier

\begin{table}[p]
\begin{center}
\resizebox{0.95\linewidth}{!}{
\begin{tabular}{P{5.0cm}|P{2.2cm}P{2.2cm}P{2.2cm}P{2.2cm}P{2.2cm}}
\hline\hline
ETF Type & Number of Funds & Investor Return & Raw Fund Return & Fund Selection Effect & Investor Timing Effect\\
\hline
\multicolumn{6}{c}{\textbf{Panel A: Long LETFs}} \\
\hline
Commodity & $10$ & $-31.1\%$ & $-26.3\%$ & $4.1\%$ & $-8.9\%$ \\
Crypto & $1$ & $44.1\%$ & $94.5\%$ & $0.0\%$ & $-50.4\%$ \\
Currency & $1$ & $-7.1\%$ & $-8.8\%$ & $0.0\%$ & $1.8\%$ \\
Levered Equity Index & $88$ & $7.8\%$ & $4.0\%$ & $12.6\%$ & $-8.8\%$ \\
Levered Single Stock & $13$ & $41.3\%$ & $46.5\%$ & $43.1\%$ & $-48.3\%$ \\
VIX & $3$ & $-78.5\%$ & $-82.6\%$ & $1.2\%$ & $2.9\%$ \\
\hline
\multicolumn{6}{c}{\textbf{Panel B: Unlevered ETFs}} \\
\hline
Bond & $9$ & $1.8\%$ & $3.0\%$ & $0.0\%$ & $-1.2\%$ \\
Cash & $2$ & $3.6\%$ & $2.0\%$ & $-0.3\%$ & $1.9\%$ \\
Commodity & $2$ & $8.4\%$ & $8.5\%$ & $-0.0\%$ & $-0.1\%$ \\
Equity Broad-Based & $14$ & $13.3\%$ & $10.8\%$ & $0.6\%$ & $1.9\%$ \\
Equity Global/International & $11$ & $6.8\%$ & $5.7\%$ & $-0.5\%$ & $1.6\%$ \\
Equity Sub-Index & $18$ & $14.3\%$ & $11.8\%$ & $0.2\%$ & $2.3\%$ \\
\hline\hline
\end{tabular}
}
\end{center}
\caption{\textbf{Investor Return Decomposition.} The table reports the investor return decomposition by ETF type for funds with at least two years of trading history. Panel A reports results for long LETFs. Panel B reports results for unlevered ETFs. Raw fund return is the equal-weighted average annualized return across funds. The fund selection effect is the difference between the investor-capital-weighted average fund return and the equal-weighted raw fund return, using investor dollar-days as weights. The investor timing effect captures whether investor capital is concentrated in funds during periods with higher or lower returns. Investor return is the sum of raw fund return, fund selection, and investor timing.}
\label{tab:investor_return_decomp_combined}
\end{table}
\FloatBarrier

\appendix
\section*{Appendix}
\addcontentsline{toc}{section}{Appendix}
\setcounter{figure}{0}
\setcounter{table}{0}
\renewcommand{\thefigure}{A\arabic{figure}}
\renewcommand{\thetable}{A\arabic{table}}

\section{Additional Results}\label{app:additional_results}

This appendix reports additional figures and robustness tables. Appendix Section~\ref{app:index_vs_stock} provides additional data contrasting single-stock LETFs with broad-based equity-index LETFs.  Appendix Section~\ref{app:vol_drag_details} gives numerical examples and the return approximation for volatility drag. Appendix Sections~\ref{app:return_robustness} and~\ref{app:flow_robustness} report additional return and flow robustness tests.

\subsection{Index and Single-Stock Products}\label{app:index_vs_stock}

\begin{figure}[h!]
\centering
\includegraphics[
height=3in,
alt={A line chart compares the monthly average daily volatility of assets tracked by leveraged equity-index ETFs and leveraged single-stock ETFs. Single-stock LETFs, introduced in the 2020s, track much more volatile assets than broad equity-index LETFs.}
]{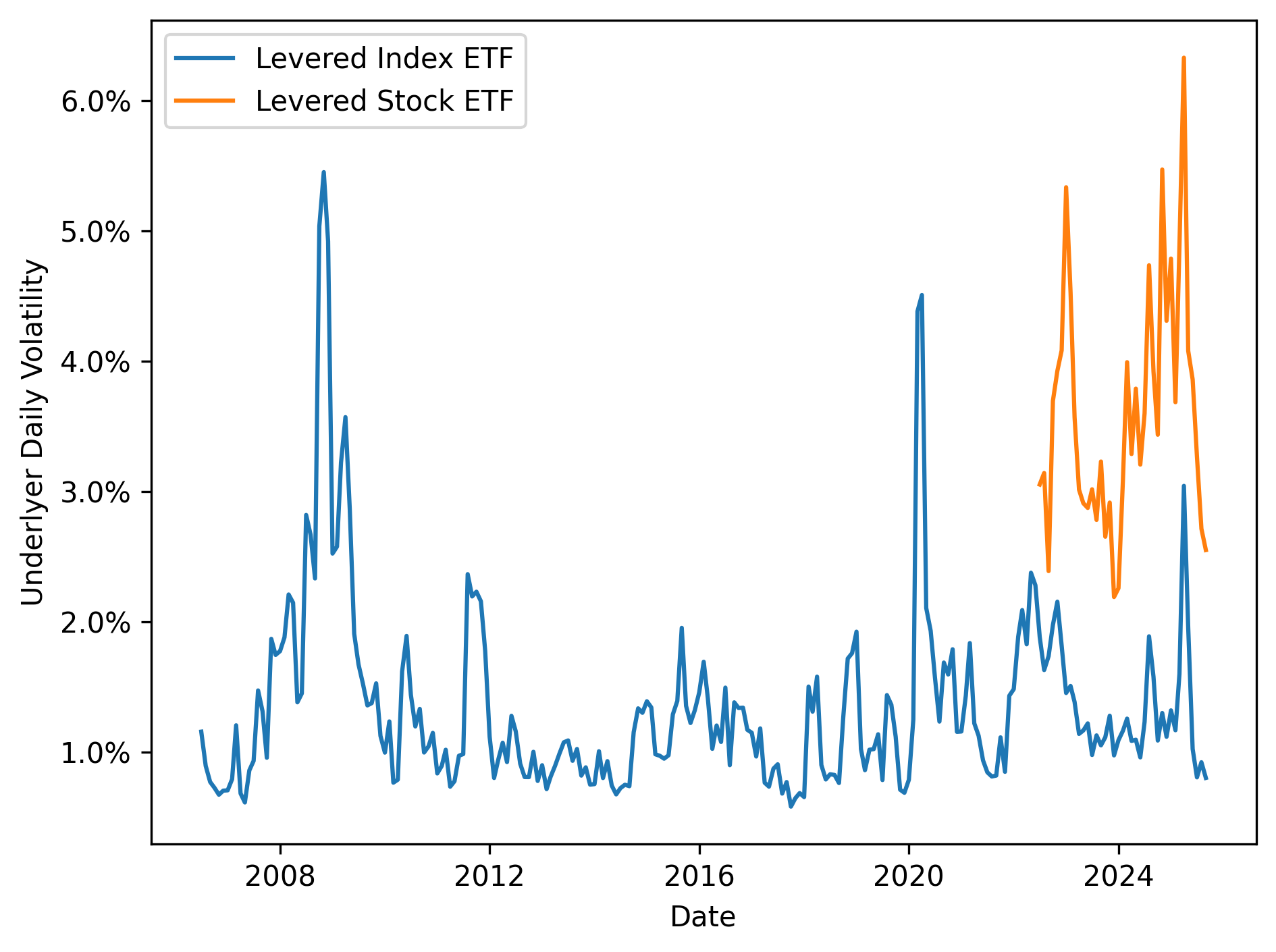}
\caption[]{\textbf{Average Daily Volatility of Assets Tracked by LETFs.} The figure reports monthly average daily volatility of assets tracked by leveraged and inverse ETFs, weighted by AUM.}
\figalt{A line chart compares the monthly average daily volatility of assets tracked by leveraged equity-index ETFs and leveraged single-stock ETFs. Single-stock LETFs track much more volatile assets than broad equity-index LETFs, especially after their introduction in the 2020s.}
\label{fig:app_idx_vs_stk_letf_vol}
\end{figure}

The main text shows that single-stock LETFs are tied to much more volatile assets than broad equity-index LETFs. Figure~\ref{fig:app_idx_vs_stk_letf_vol} reports the same comparison over time. Panel B of Table~\ref{tab:breakeven_vol_drag} reports annual averages. In 2025, the asset-weighted average daily volatility of assets tracked by leveraged single-stock ETFs was 3.71\%, compared with 1.35\% for leveraged equity-index ETFs.

Because the log volatility-drag term scales with the square of volatility, this difference is economically large. Holding leverage fixed, moving from 1.35\% daily volatility to 3.71\% daily volatility increases the log volatility-drag term by about 7.6 times. The final two columns of Panel B of Table~\ref{tab:breakeven_vol_drag} translate this difference into annualized volatility drag for a $2\times$ long LETF.

Appendix Table~\ref{tab:app_rank_aum_regressions_vs_vol} provides a rank-based robustness check for the AUM regressions in Table~\ref{tab:single_stock_launch_aum}.  The results are qualitatively similar: underlying asset volatility is positively related to LETF AUM.  The coefficients around $0.4$ indicate that a 10-percentile-point increase in the underlying stock's volatility rank is associated with a 4-percentile-point higher AUM rank among single-stock LETFs, holding market-capitalization rank fixed and, in column (1), fund age fixed.

\begin{table}\begin{center}
\begin{tabular}{P{6.0cm}|P{2.2cm}P{2.2cm}}
\hline\hline
  & (1) & (2)\\
\hline
Underlying Volatility Rank & $0.41$ & $0.38$ \\
 & $(3.74)$ & $(3.29)$ \\
Underlying Market Cap Rank & $-0.51$ & $0.36$ \\
 & $(-1.06)$ & $(5.46)$ \\
Log Number of Fund Days & $0.14$ &  \\
 & $(1.77)$ &  \\
\hline
Observations & $40$ & $40$ \\
Adjusted $R^2$ & $0.87$ & $0.87$ \\
\hline\hline\end{tabular}
\end{center}\caption[] { \textbf{Single-Stock LETF AUM Rank Regressions.} The table reports fund-level OLS regressions for long leveraged single-stock ETFs whose underlying stocks are in the top-1,000-stock universe by market capitalization and that have at least 252 fund-day observations. The dependent variable is the fund's percentile rank in average daily AUM among single-stock LETFs over the fund's sample period. The main explanatory variables are the underlying stock's volatility percentile rank and market-capitalization percentile rank, each computed relative to the top 1,000 stocks by market capitalization and averaged over the fund's sample period. Column (1) controls for fund age using the log number of fund trading days, while Column (2) omits this control. All percentile ranks are increasing in the underlying variable. Regressions are estimated by OLS with heteroskedasticity-robust standard errors. t-statistics are reported in parentheses.}\label{tab:app_rank_aum_regressions_vs_vol}\end{table}

\FloatBarrier

\subsection{Inverse LETFs}\label{app:inverse_letfs}

\begin{figure}[t]
\centering
\includegraphics[
width=0.82\linewidth,
alt={A bar chart shows annual investor PnL and average daily assets under management for inverse equity-index and single-stock LETFs. Average AUM is positive in every year, while investor PnL is negative in most years, indicating persistent investor losses in inverse LETFs despite meaningful assets under management.}
]{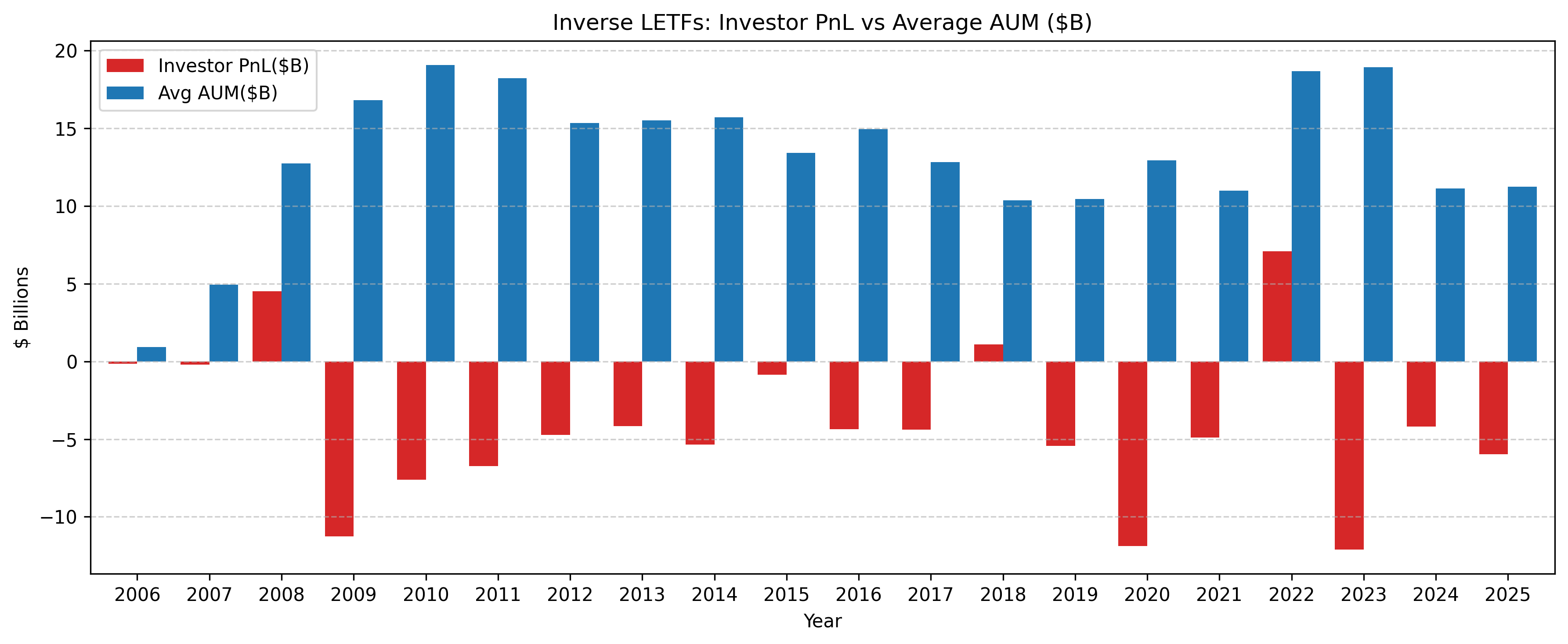}
\caption{\textbf{Investor PnL in Inverse LETFs.} The figure reports annual investor PnL and average daily assets under management for inverse equity-index and single-stock LETFs.}
\figalt{A bar chart shows annual investor PnL and average daily assets under management for inverse equity-index and single-stock LETFs. Average AUM is positive in every year, while investor PnL is negative in most years, indicating persistent investor losses in inverse LETFs despite meaningful assets under management.}
\label{fig:letf_inverse_pnl1}
\end{figure}

Figure~\ref{fig:letf_inverse_pnl1} shows investor PnL and AUM in inverse LETFs by year. The figure provides additional detail on the persistent losses in inverse LETFs discussed in the main text.

\FloatBarrier

\subsection{Cross-Sectional Return Robustness}\label{app:return_robustness}

Tables~\ref{tab:app_letf_reg2} and~\ref{tab:app_letf_reg3} report cross-sectional return regressions for long LETFs. The results support the main return decomposition in the paper. Predicted volatility drag is strongly negatively related to average log returns, while returns on the asset tracked by the product are strongly positively related to LETF returns.

\begin{table}[t]
\begin{center}
\resizebox{0.95\linewidth}{!}{
\begin{tabular}{P{3.2cm}|P{2.1cm}P{2.1cm}P{2.1cm}P{2.3cm}P{2.3cm}}
\hline\hline
Group & Number of Levered ETFs & Number Outperforming Underlying & Pct. Outperforming & Mean Underlying Annualized Log Return & Mean LETF Annualized Log Return\\
\hline
\multicolumn{6}{c}{\textbf{Panel A: At Least 3.0 Years of Trading History, by ETF Type}} \\
\hline
Commodity & $8$ & $1$ & $12.5\%$ & $-9.7\%$ & $-54.8\%$ \\
Levered Equity Index & $46$ & $20$ & $43.5\%$ & $9.4\%$ & $2.9\%$ \\
Levered Single Stock & $7$ & $3$ & $42.9\%$ & $21.6\%$ & $16.4\%$ \\
\hline
\multicolumn{6}{c}{\textbf{Panel B: At Least 3.0 Years of Trading History, by Underlying Daily Volatility}} \\
\hline
1. 0\% - 1.5\% & $27$ & $19$ & $70.4\%$ & $9.8\%$ & $12.9\%$ \\
2. 1.5\% - 2\% & $17$ & $5$ & $29.4\%$ & $10.7\%$ & $5.1\%$ \\
3. 2\% - 3\% & $14$ & $0$ & $0.0\%$ & $2.3\%$ & $-39.5\%$ \\
4. $>$ 3\% & $3$ & $0$ & $0.0\%$ & $8.0\%$ & $-23.9\%$ \\
\hline
\multicolumn{6}{c}{\textbf{Panel C: At Least 1.5 Years of Trading History, by ETF Type}} \\
\hline
Commodity & $8$ & $1$ & $12.5\%$ & $-9.7\%$ & $-54.8\%$ \\
Levered Equity Index & $47$ & $21$ & $44.7\%$ & $9.6\%$ & $3.6\%$ \\
Levered Single Stock & $18$ & $8$ & $44.4\%$ & $32.1\%$ & $31.4\%$ \\
\hline
\multicolumn{6}{c}{\textbf{Panel D: At Least 1.5 Years of Trading History, by Underlying Daily Volatility}} \\
\hline
1. 0\% - 1.5\% & $29$ & $20$ & $69.0\%$ & $10.4\%$ & $13.7\%$ \\
2. 1.5\% - 2\% & $19$ & $5$ & $26.3\%$ & $11.3\%$ & $6.1\%$ \\
3. 2\% - 3\% & $16$ & $2$ & $12.5\%$ & $7.9\%$ & $-27.2\%$ \\
4. $>$ 3\% & $9$ & $3$ & $33.3\%$ & $34.2\%$ & $24.0\%$ \\
\hline\hline
\end{tabular}
}
\end{center}
\caption{\textbf{LETF Outperformance Relative to the Underlying Asset.} The table reports the number of LETFs and the fraction outperforming their underlying asset. The sample includes long LETFs where the underlying asset is a stock index, stock, or commodity. Panels A and B include LETFs with at least 3.0 years of trading history. Panels C and D include LETFs with at least 1.5 years of trading history. Panels A and C report results by ETF type, while Panels B and D report results by underlying daily volatility bucket.}
\label{tab:etfvunder_combined}
\end{table}

\begin{table}\begin{center}\resizebox{0.95\linewidth}{!}{\begin{tabular}{P{3.2cm}|P{2.2cm}P{2.2cm}P{2.2cm}P{2.2cm}P{2.2cm}P{2.2cm}P{2.2cm}P{2.2cm}}
\hline\hline
 & (1) & (2) & (3) & (4) & (5) & (6) & (7) & (8)\\
\hline
Intercept & $0.000$ & $0.000$ & $0.000$ & $0.000$ & $0.000$ & $0.000$ & $0.000$ & $0.000$ \\
 & $(0.92)$ & $(1.01)$ & $(3.41)$ & $(3.25)$ & $(0.69)$ & $(0.53)$ & $(2.77)$ & $(2.67)$ \\
PredVolDrag & $-0.895$ & $-0.892$ & $-0.867$ & $-0.855$ & $-0.921$ & $-0.914$ & $-0.867$ & $-0.860$ \\
 & $(-23.72)$ & $(-27.40)$ & $(-29.14)$ & $(-32.88)$ & $(-30.07)$ & $(-31.85)$ & $(-26.57)$ & $(-27.95)$ \\
TrackedAssetRet & $0.938$ & $0.942$ & $0.951$ & $0.957$ & $0.936$ & $0.939$ & $0.955$ & $0.958$ \\
 & $(51.57)$ & $(53.54)$ & $(59.90)$ & $(63.58)$ & $(43.42)$ & $(44.56)$ & $(50.04)$ & $(52.82)$ \\
DailyDrag & $1.777$ & $1.728$ & $1.728$ & $1.671$ & $1.756$ & $1.707$ & $1.711$ & $1.664$ \\
 & $(16.52)$ & $(11.81)$ & $(16.85)$ & $(12.60)$ & $(15.87)$ & $(11.78)$ & $(17.34)$ & $(13.20)$ \\
RFFunding & & & $1.359$ & $1.611$ & & & $1.379$ & $1.548$ \\
 & & & $(5.18)$ & $(5.43)$ & & & $(3.46)$ & $(3.92)$ \\
AutoCorr1 & & & & & $-0.001$ & $-0.001$ & $0.000$ & $-0.000$ \\
 & & & & & $(-3.47)$ & $(-2.68)$ & $(0.08)$ & $(-0.22)$ \\
AutoCorr21 & & & & & $-0.000$ & $-0.000$ & $-0.000$ & $-0.000$ \\
 & & & & & $(-0.05)$ & $(-0.12)$ & $(-0.73)$ & $(-0.45)$ \\
\hline
Observations & $51$ & $51$ & $51$ & $51$ & $51$ & $51$ & $51$ & $51$ \\
Adjusted $R^2$ & $0.998$ & $0.998$ & $0.999$ & $0.999$ & $0.998$ & $0.999$ & $0.999$ & $0.999$ \\
\hline
Weight & Equal Weight & NDays & Equal Weight & NDays & Equal Weight & NDays & Equal Weight & NDays \\
\hline\hline\end{tabular}
}\end{center}\caption[] { \textbf{Cross Sectional Return Regression: January 2011 to February 2020.} The table presents a cross-sectional regression of the average log return of long LETFs over the period 2011-01-03 to 2020-02-28 on predicted volatility drag, the return of the asset tracked by the fund, the daily drag term, risk-free funding, and autocorrelation controls. Each regression is run with two weighting schemes: one in which observations are equally weighted and one in which they are weighted by the number of days during the sample period for which the ETF has valid data. T-statistics computed using robust HC0 standard errors \citep{standarderrors} are reported in parentheses below the point estimates.}\label{tab:app_letf_reg2}\end{table}

\begin{table}\begin{center}\resizebox{0.95\linewidth}{!}{\begin{tabular}{P{3.2cm}|P{2.2cm}P{2.2cm}P{2.2cm}P{2.2cm}P{2.2cm}P{2.2cm}P{2.2cm}P{2.2cm}}
\hline\hline
 & (1) & (2) & (3) & (4) & (5) & (6) & (7) & (8)\\
\hline
Intercept & $-0.000$ & $-0.000$ & $0.000$ & $0.000$ & $-0.000$ & $-0.000$ & $0.000$ & $0.000$ \\
 & $(-3.97)$ & $(-4.09)$ & $(1.51)$ & $(1.47)$ & $(-3.75)$ & $(-3.83)$ & $(1.69)$ & $(1.65)$ \\
PredVolDrag & $-1.002$ & $-1.006$ & $-0.921$ & $-0.921$ & $-1.006$ & $-1.009$ & $-0.924$ & $-0.924$ \\
 & $(-28.28)$ & $(-27.10)$ & $(-28.11)$ & $(-27.63)$ & $(-28.39)$ & $(-26.73)$ & $(-29.25)$ & $(-28.16)$ \\
TrackedAssetRet & $0.945$ & $0.945$ & $0.969$ & $0.969$ & $0.941$ & $0.942$ & $0.965$ & $0.965$ \\
 & $(38.15)$ & $(39.29)$ & $(73.52)$ & $(75.82)$ & $(34.48)$ & $(35.40)$ & $(66.80)$ & $(67.91)$ \\
DailyDrag & $0.780$ & $0.742$ & $0.772$ & $0.753$ & $0.774$ & $0.739$ & $0.765$ & $0.748$ \\
 & $(2.64)$ & $(2.40)$ & $(3.32)$ & $(3.18)$ & $(2.73)$ & $(2.51)$ & $(3.74)$ & $(3.58)$ \\
RFFunding & & & $1.420$ & $1.419$ & & & $1.448$ & $1.446$ \\
 & & & $(8.77)$ & $(9.03)$ & & & $(9.47)$ & $(9.66)$ \\
AutoCorr1 & & & & & $-0.000$ & $-0.000$ & $-0.000$ & $-0.000$ \\
 & & & & & $(-0.89)$ & $(-0.83)$ & $(-1.82)$ & $(-1.78)$ \\
AutoCorr21 & & & & & $-0.000$ & $-0.000$ & $-0.001$ & $-0.001$ \\
 & & & & & $(-0.95)$ & $(-0.87)$ & $(-1.84)$ & $(-1.77)$ \\
\hline
Observations & $46$ & $46$ & $46$ & $46$ & $46$ & $46$ & $46$ & $46$ \\
Adjusted $R^2$ & $0.992$ & $0.992$ & $0.996$ & $0.996$ & $0.992$ & $0.992$ & $0.997$ & $0.997$ \\
\hline
Weight & Equal Weight & NDays & Equal Weight & NDays & Equal Weight & NDays & Equal Weight & NDays \\
\hline\hline\end{tabular}
}\end{center}\caption[] {  \textbf{Cross Sectional Return Regression: February 2020 to September 2025.} The table presents a cross-sectional regression of the average log return of long LETFs over the period 2020-03-02 to 2025-09-30 on predicted volatility drag, the return of the asset tracked by the fund, the daily drag term, risk-free funding, and autocorrelation controls. Each regression is run with two weighting schemes: one in which observations are equally weighted and one in which they are weighted by the number of days during the sample period for which the ETF has valid data. T-statistics computed using robust HC0 standard errors \citep{standarderrors} are reported in parentheses below the point estimates.}\label{tab:app_letf_reg3}\end{table}

\subsection{Flows and Turnover}\label{app:flow_robustness}

Table~\ref{tab:app_flow_equal_weight} repeats the main flow regression with equal weights rather than AUM weights. The results are similar: long LETF flows are negatively related to recent returns.

\begin{table}\begin{center}\resizebox{0.90\linewidth}{!}{\begin{tabular}{P{4.4cm}|P{2.2cm}P{2.2cm}P{2.2cm}P{2.2cm}}
\hline\hline
 & (1) & (2) & (3) & (4)\\
\hline
Lagged 1-Day Return & $-1.78$ & $-1.95$ & $-2.46$ & $-0.88$ \\
 & $(-29.49)$ & $(-10.51)$ & $(-28.64)$ & $(-6.31)$ \\
Lagged 5-Day Return & $-1.22$ & $-1.82$ & $-1.47$ & $-0.78$ \\
 & $(-42.23)$ & $(-20.65)$ & $(-35.40)$ & $(-12.37)$ \\
Lagged 21-Day Return & $-0.21$ & $-0.42$ & $-0.23$ & $-0.16$ \\
 & $(-14.75)$ & $(-8.55)$ & $(-11.01)$ & $(-5.01)$ \\
Lagged 63-Day Return & $-0.04$ & $0.03$ & $-0.07$ & $-0.07$ \\
 & $(-4.26)$ & $(0.68)$ & $(-5.78)$ & $(-3.51)$ \\
Lagged 126-Day Return & $0.01$ & $-0.07$ & $0.05$ & $0.13$ \\
 & $(1.98)$ & $(-3.26)$ & $(6.22)$ & $(8.81)$ \\
Intercept & $0.00$ & $-0.00$ & $0.00$ & $0.00$ \\
 & $(0.34)$ & $(-0.20)$ & $(0.58)$ & $(0.55)$ \\
\hline
Observations & $323,786$ & $30,119$ & $161,295$ & $38,758$ \\
Adjusted $R^2$ & $0.03$ & $0.06$ & $0.04$ & $0.01$ \\
Period & Entire Sample & 2006 to 2011 & 2012 to 2020 & 2024 and 2025 \\
\hline\hline\end{tabular}
}\end{center}\caption[] { \textbf{Regression of Normalized Flows on Past Returns.} For long LETFs in our sample, we regress normalized daily flows on lagged returns. Flows are winsorized at the 2.5\% and 97.5\% levels within fund. The independent variables are lagged LETF returns, winsorized at the 2.5\% and 97.5\% levels. We run separate regressions for each time period, and cluster standard errors by date to account for cross-sectional correlation in flows across ETFs on the same day. Observations are weighted equally during the period.}\label{tab:app_flow_equal_weight}\end{table}

One possible concern is that the flow results are driven by market makers hedging gamma and creating shares after LETF underperformance. This explanation would require market makers to be systematically short gamma. Table~\ref{tab:app_flow_month_in_quarter} addresses this possibility by estimating the flow regressions separately by month within the quarter. If gamma hedging around option expiration were driving the results, the effect should be stronger in the third month of the quarter. The estimates are similar across months, suggesting that option-expiration-related hedging is not driving the main flow results.

\begin{table}\begin{center}\resizebox{0.90\linewidth}{!}{\begin{tabular}{P{4.4cm}|P{2.2cm}P{2.2cm}P{2.2cm}}
\hline\hline
 & (1) & (2) & (3)\\
\hline
Lagged 1-Day Return & $-3.39$ & $-3.47$ & $-2.55$ \\
 & $(-13.40)$ & $(-14.02)$ & $(-10.59)$ \\
Lagged 5-Day Return & $-2.00$ & $-1.81$ & $-2.08$ \\
 & $(-16.59)$ & $(-15.16)$ & $(-18.05)$ \\
Lagged 21-Day Return & $-0.24$ & $-0.25$ & $-0.49$ \\
 & $(-4.39)$ & $(-4.32)$ & $(-8.09)$ \\
Lagged 63-Day Return & $-0.12$ & $-0.06$ & $0.06$ \\
 & $(-3.00)$ & $(-1.34)$ & $(1.51)$ \\
Lagged 126-Day Return & $0.06$ & $-0.01$ & $0.03$ \\
 & $(2.75)$ & $(-0.57)$ & $(1.29)$ \\
Intercept & $-0.00$ & $0.01$ & $0.03$ \\
 & $(-0.37)$ & $(1.08)$ & $(4.41)$ \\
\hline
Observations & $104,107$ & $102,730$ & $105,614$ \\
Adjusted $R^2$ & $0.06$ & $0.06$ & $0.06$ \\
Month in Quarter & 1st Month in Qtr & 2nd Month in Qtr & 3rd Month in Qtr \\
\hline\hline\end{tabular}
}\end{center}\caption[] { \textbf{Regression of Normalized Flows on Lagged Returns.} For long LETFs in our sample, we regress normalized daily flows on lagged returns. Flows are winsorized at the 2.5\% and 97.5\% levels within fund. The independent variables are lagged LETF returns, winsorized at the 2.5\% and 97.5\% levels. We run separate regressions for each time period, and cluster standard errors by date to account for cross-sectional correlation in flows across ETFs on the same day. Observations are weighted by each fund's total AUM over the months included in the corresponding month-in-quarter regression.}\label{tab:app_flow_month_in_quarter}\end{table}

Table~\ref{tab:app_unlevered_future_flows} repeats the future-return flow specification for unlevered ETFs. Consistent with the main results, flows do not have a stable positive relation with future returns.

\begin{table}\begin{center}\resizebox{0.90\linewidth}{!}{\begin{tabular}{P{4.4cm}|P{2.2cm}P{2.2cm}P{2.2cm}P{2.2cm}}
\hline\hline
 & (1) & (2) & (3) & (4)\\
\hline
Future 1-Day Return & $0.16$ & $-0.13$ & $0.72$ & $-1.19$ \\
 & $(0.47)$ & $(-0.24)$ & $(1.13)$ & $(-0.86)$ \\
Future 5-Day Return & $-0.12$ & $-0.28$ & $-0.61$ & $0.09$ \\
 & $(-0.69)$ & $(-0.91)$ & $(-1.92)$ & $(0.14)$ \\
Future 21-Day Return & $-0.05$ & $-0.06$ & $-0.62$ & $0.26$ \\
 & $(-0.57)$ & $(-0.37)$ & $(-3.70)$ & $(0.65)$ \\
Future 63-Day Return & $-0.26$ & $-0.41$ & $0.19$ & $-0.65$ \\
 & $(-5.03)$ & $(-5.23)$ & $(2.22)$ & $(-2.80)$ \\
Lagged 1-Day Return & $2.06$ & $2.51$ & $4.27$ & $-0.35$ \\
 & $(5.93)$ & $(4.36)$ & $(6.70)$ & $(-0.26)$ \\
Lagged 5-Day Return & $1.70$ & $1.82$ & $2.24$ & $2.17$ \\
 & $(9.58)$ & $(5.72)$ & $(7.15)$ & $(3.60)$ \\
Lagged 21-Day Return & $0.20$ & $0.60$ & $0.68$ & $-0.47$ \\
 & $(2.23)$ & $(3.63)$ & $(4.31)$ & $(-1.61)$ \\
Lagged 63-Day Return & $0.30$ & $-0.40$ & $0.55$ & $0.39$ \\
 & $(6.32)$ & $(-4.63)$ & $(6.69)$ & $(2.14)$ \\
Intercept & $-0.01$ & $0.00$ & $-0.03$ & $-0.00$ \\
 & $(-3.02)$ & $(0.24)$ & $(-5.42)$ & $(-0.06)$ \\
\hline
Observations & $225,813$ & $43,016$ & $118,984$ & $19,817$ \\
Adjusted $R^2$ & $0.00$ & $0.01$ & $0.01$ & $0.00$ \\
Period & Entire Sample & 2006 to 2011 & 2012 to 2020 & 2024 and 2025 \\
\hline\hline\end{tabular}
}\end{center}\caption[] { \textbf{Regression of Unlevered ETF Flows on Lagged and Future Returns.} For unlevered ETFs in our sample, we regress normalized daily flows on lagged and future returns. Flows are winsorized at the 2.5\% and 97.5\% levels within fund. The independent variables are the winsorized lagged and future returns of the unlevered ETFs, with winsorization at the 2.5\% and 97.5\% level. We run separate regressions for each time period, and cluster standard errors by date to account for cross-sectional correlation in flows across ETFs on the same day. Observations are weighted by each fund's total AUM over the months included in the corresponding regression.}\label{tab:app_unlevered_future_flows}\end{table}

As a final robustness check, Table~\ref{tab:app_pooledFlowRegression} estimates a pooled regression of daily normalized flows on the lagged one-day return of the underlying asset, interacted with indicators for long LETFs, inverse LETFs, and unlevered ETFs. The pattern is stark. For long LETFs, the coefficients are negative in every period, indicating that investors buy after the underlying asset falls and sell after it rises. Because long LETFs move in the same direction as their underlying assets, these flows are contrarian with respect to the fund's own recent return.

Inverse LETFs show the corresponding pattern through the opposite side of the trade. The coefficients are positive in every period, meaning that investors buy inverse products after the underlying asset rises and sell after it falls. Since inverse LETFs gain when the underlying asset falls, these flows are consistent with investors increasing inverse exposure after increases in the underlying asset. In this sense, both long and inverse LETF flows point toward reversal-oriented trading: investors buy long LETFs after declines in the underlying and buy inverse LETFs after increases in the underlying. Unlevered ETFs look different. Their positive coefficients indicate that flows move in the same direction as past returns. The pooled evidence therefore reinforces the main result: LETF investors trade against recent underlying returns, while unlevered ETF investors trade with them.

\begin{table}\begin{center}\resizebox{0.90\linewidth}{!}{\begin{tabular}{P{8.0cm}|P{2.2cm}P{2.2cm}P{2.2cm}P{2.2cm}}
\hline\hline
  & (1) & (2) & (3) & (4)\\
\hline
Lagged Underlying Return $\times$ Long & $-4.26$ & $-10.52$ & $-6.63$ & $-2.49$ \\
 & $(-18.64)$ & $(-15.19)$ & $(-13.20)$ & $(-7.57)$ \\
Lagged Underlying Return $\times$ Inverse & $4.07$ & $3.84$ & $3.51$ & $5.36$ \\
 & $(15.62)$ & $(8.46)$ & $(7.10)$ & $(12.78)$ \\
Lagged Underlying Return $\times$ Unlevered & $3.80$ & $2.57$ & $5.40$ & $3.76$ \\
 & $(20.27)$ & $(8.74)$ & $(15.09)$ & $(5.37)$ \\
Intercept & $-0.00$ & $-0.00$ & $-0.00$ & $-0.00$ \\
 & $(-1.49)$ & $(-0.80)$ & $(-0.45)$ & $(-0.41)$ \\
\hline
Observations & $438,987$ & $74,362$ & $220,931$ & $56,230$ \\
Adjusted $R^2$ & $0.00$ & $0.01$ & $0.00$ & $0.01$ \\
Period & Entire Sample & 2006 to 2011 & 2012 to 2020 & 2024 and 2025 \\
\hline\hline\end{tabular}
}\end{center}\caption[] { \textbf{Flow Responses to Lagged Underlying Returns.} The table reports pooled daily flow regressions for unlevered ETFs, long LETFs, and inverse LETFs. The dependent variable is daily normalized fund flow. The independent variables are lagged one-day underlying returns interacted with indicators for each fund type. For unlevered ETFs, the underlying asset is the ETF itself. For long LETFs, a positive coefficient means that investors buy more after the underlying asset rises. For inverse LETFs, a positive coefficient means that investors buy more after the underlying asset rises, even though the inverse LETF itself falls when the underlying asset rises. Thus, a positive inverse-LETF coefficient means that investors buy inverse products after the inverse products themselves have fallen. Columns report estimates for the full sample and for three subperiods. The sample is restricted to funds with at least 252 daily observations in the relevant period. Standard errors are clustered by date. t-statistics are reported in parentheses.}\label{tab:app_pooledFlowRegression}\end{table}

\subsection{Volatility Drag Details}\label{app:vol_drag_details}

The multi-day return of an LETF can differ substantially from the multi-day return of the asset the product tracks compounded by the leverage factor. Suppose the asset tracked by an ETF rises from \$100 to \$105 on day 1 and falls back to \$100 on day 2. The underlying asset has returns of 5\% and -4.76\%, so its two-day compound return is zero. A $2\times$ long LETF has daily returns of 10\% and -9.52\%, for a two-day return of -0.48\%. A $1\times$ inverse product has the same two-day return in this example. A $3\times$ long LETF and a $2\times$ inverse LETF both lose 1.43\%. If the asset instead rises from \$100 to \$110 and then falls back to \$100, the $2\times$ long LETF and the $1\times$ inverse product lose 1.82\%, while the $3\times$ long LETF and $2\times$ inverse product lose 5.45\%.

The same mechanics also create a benefit in trending markets. If the asset rises from \$100 to \$110 to \$121, the underlying asset earns 10\% per day and has a two-day return of 21\%. A $2\times$ LETF earns 20\% per day and has a two-day return of 44\%. This is larger than twice the underlying asset's two-day return because, before fees and financing costs, LETFs \textbf{compound} daily levered returns rather than \textbf{multiply} the underlying asset's multi-period return.

Let $R_u$ be the return of the asset tracked by the LETF and let $R_F$ be the daily risk-free rate. Ignoring fees and financing spreads, a levered ETF with leverage ratio $L$ has daily return $R_L=L R_u-(L-1)R_F$. Following \citet{cheng2009dynamics}, the expected log return of a LETF is approximately
\begin{equation} \label{eq:app_voldrag}
E[\log(1+R_L)] = L E[\log(1+R_u)] - (L-1) R_F - \frac{L(L-1)}{2}\sigma_u^2.
\end{equation}
The last term is volatility drag. It is proportional to the square of the volatility of the asset the product tracks. If volatility triples, the volatility-drag term increases by a factor of nine. It is also proportional to $L(L-1)$, so a $3\times$ LETF has about three times the volatility drag of a $2\times$ LETF for the same asset and horizon.

Two points about this term are worth emphasizing. First, volatility drag is not a special consequence of the one-day reset frequency. To the leading order captured by Equation~\ref{eq:app_voldrag}, the same drag appears with any fixed rebalancing interval. For example, if the fund rebalances every $k$ days rather than every day, the variance of the $k$-day underlying return is approximately $k\sigma_u^2$ when daily returns are independent. At the same time, the number of rebalancing periods over a fixed horizon falls by the same factor, from $N$ to $N/k$. The product of the number of periods and the per-period variance is therefore approximately unchanged:
\[
  \frac{N}{k}\cdot k\sigma_u^2 = N\sigma_u^2.
\]
Thus, changing the reset interval changes higher-order terms and implementation details, but it does not eliminate the leading volatility-drag term.

Second, volatility drag does not require negative autocorrelation or buy-high/sell-low predictability in the underlying asset. Equation~\ref{eq:app_voldrag} holds even when underlying returns are independently and identically distributed. Autocorrelation can affect realized multi-period outcomes because it changes the path of returns and, for longer reset intervals, the variance of reset-period returns. But the basic volatility-drag term comes from compounding leveraged returns and the concavity of the log function, not from predictable reversals in the underlying asset.

\subsection{Counterfactual Underlying Investment Details}
\label{app:counterfactual_details}

Let $B_{i,t}$ denote the value of the counterfactual investment in the underlying asset for LETF $i$. Before flows on day $t$, the counterfactual account evolves as
\[
  B^{pre}_{i,t}
  =
  B_{i,t-1}(1+R^{u}_{i,t}),
\]
where $R^{u}_{i,t}$ is the return on the underlying asset. If the LETF has a positive flow, we add that dollar flow to the counterfactual account. If the LETF has a negative flow, we compute the redemption fraction from the LETF and apply the same fraction to the counterfactual account:
\[
B_{i,t}
=
\begin{cases}
B^{pre}_{i,t}+\text{Flow}_{i,t}, & \text{if } \text{Flow}_{i,t}\geq 0, \\
B^{pre}_{i,t}\left(1-\frac{-\text{Flow}_{i,t}}{A^{pre}_{i,t}}\right), & \text{if } \text{Flow}_{i,t}<0,
\end{cases}
\]
where $A^{pre}_{i,t}$ is the pre-flow value of assets in the LETF after day-$t$ returns.

\subsection{Return Decomposition Details}
\label{app:return_decomp_details}

Let $A_{i,t}$ denote total assets in fund $i$ after flows on day $t$, and let $K_{i,t}$ denote investor capital remaining in the fund after flows on day $t$. Inflows increase investor capital dollar-for-dollar. Outflows reduce investor capital pro rata by the same redemption rate as fund assets.

Daily investor PnL is
\[
  \text{PnL}_{i,t}
  =
  A_{i,t-1}R_{i,t}.
\]
After returns are realized, pre-flow assets are
\[
  A^{pre}_{i,t}
  =
  A_{i,t-1}(1+R_{i,t}).
\]
Investor capital evolves as
\[
K_{i,t}
=
\begin{cases}
K_{i,t-1}+\text{Flow}_{i,t}, & \text{if } \text{Flow}_{i,t}\geq 0, \\
K_{i,t-1}\left(1-\frac{-\text{Flow}_{i,t}}{A^{pre}_{i,t}}\right), & \text{if } \text{Flow}_{i,t}<0.
\end{cases}
\]
This treats withdrawals as removing a proportional share of remaining investor capital. For example, if a fund doubles in value and investors redeem half of the fund, the redemption removes half of the remaining investor capital, not the full dollar amount of the withdrawal.

For each fund, we compute the annualized time-weighted return:
\[
R^{TW}_{i}
=
\exp\left(
252 \times \frac{1}{N_i}\sum_t \log(1+R_{i,t})
\right)-1,
\]
where $N_i$ is the number of fund-day observations. We then compute the investor-capital-weighted annualized return:
\[
R^{ICW}_{i}
=
\exp\left(
252 \times
\frac{\sum_t K_{i,t-1}\log(1+R_{i,t})}
{\sum_t K_{i,t-1}}
\right)-1.
\]
The within-fund timing effect is
\[
  \text{Timing Effect}_{i}
  =
  R^{ICW}_{i}-R^{TW}_{i}.
\]

\newpage
\clearpage

\bibliographystyle{apalike}
\bibliography{bib}

\end{document}